\documentclass[reprint,amsmath,amssymb,textcomp,aps]{revtex4-1}
\usepackage{amsmath}
\usepackage{amssymb}
\usepackage{textcomp}
\usepackage{graphicx}
\usepackage{dcolumn}
\usepackage{bm}
\usepackage[pdftex]{color}
\usepackage[pdftex]{hyperref}
\hypersetup{
             final,
             colorlinks=true,
             linkcolor=blue,
             citecolor=blue,
             urlcolor=blue
}

\begin{document}

\preprint{APS/123-QED}

\title{Phase diagram of electron-hole liquid in monolayer heterostructures based on transition metal dichalcogenides}

\author{P.~L.~Pekh}
\email{pavel.pekh@phystech.edu}
\affiliation{P.~N.~Lebedev Physical Institute, Russian Academy of Sciences, Leninski\u{\i} pr. 53, 119991 Moscow, Russia}

\author{P.~V.~Ratnikov}
\email{ratnikov@lpi.ru}
\affiliation{A.~M.~Prokhorov General Physics Institute, Russian Academy of Sciences, ul. Vavilova 38, 119991 Moscow, Russia}

\author{A.~P.~Silin}
\email{a.p.silin@mail.ru}
\affiliation{P.~N.~Lebedev Physical Institute, Russian Academy of Sciences, Leninski\u{\i} pr. 53, 119991 Moscow, Russia\\
Moscow Institute of Physics and Technology (National Research University), Dolgoprudnyi, 141700 Moscow region, Russia}

\date{\today}

\begin{abstract}
In the last few years, interest in monomolecular layers of transition metal dichalcogenides (TMDs) has been driven by their unusual electronic and optical properties, which are very attractive for designing functional elements of new generation nanoelectronics. Recently, experimental evidence has appeared for the formation of a high-temperature strongly bound electron-hole liquid (EHL) in TMD monolayers. Strong binding of charge carriers is due to a significant decrease in the screening of the Coulomb interaction in monolayer heterostructures. In this work, the gas--liquid phase transition in a system of electrons and holes in quasi-two-dimensional heterostructures based on TMDs is considered and the critical parameters of such a transition are calculated. The metal-insulator transition is described and the phase diagram for such heterostructures is constructed.
\end{abstract}

\pacs{71.35.−y, 73.21.Fg, 73.90.+f}

\keywords{2D materials, heterostructures, dichalcogenides, electron-hole liquid, excitons}
\maketitle

\section{\label{s1}Introduction}
The last fifteen years have been marked by intensive research of two-dimensional (2D) carbon material, graphene, as the basis for new generation nanoelectronics \cite{RS2018}. Extensive activities have arisen to study 2D materials such as monolayers of hexagonal boron nitride, black phosphorus, transition metal dichalcogenides (TMDs), and many other compounds \cite{Miro2014}. A promising direction of this activity has become the creation and study of vertical (van der Waals) heterostructures, in which various 2D materials  are combined in a given sequence \cite{Geim2013}.

Monomolecular layers of TMDs fill a special place in the current studies of 2D materials. They have the structure of a sandwich with a layer of transition metal atoms $M$, enclosed between layers of chalcogen atoms $X$, and are described by the chemical formula $MX_2$. The most studied representatives of this class of substances are compounds with metal atoms of group VI ($M$ = Mo, W) and S, Se, Te as a chalcogen. Bulk layered TMDs, such as MoS$_2$, WS$_2$, MoSe$_2$, WSe$_2$, have an indirect bandgap $E_g\sim1$ eV \cite{Bulaevskii1975, Bulaevskii1976}, and their monomolecular layers become direct-gap semiconductors with $E_g$ about 2 eV \cite{Durnev2018}.

In TMD monolayers, it is possible to achieve valley-selective excitation of charge carriers by a circularly polarized electromagnetic wave \cite{Zeng, Mak1, XiaoD, CaoT, Jones}. This prompted researchers to develop a new type of nanoelectronic devices, valleytronics, where a selective transfer of charge carriers with a certain valley index would take place \cite{RS2018, WangQH, Mak2, Mak3}.

The optical properties of monomolecular layers of TMDs are largely determined by excitons and trions. The binding energy of the exciton $E_x$ in TMD is hundreds of meV, and of the trion is tens of meV \cite{Durnev2018}. The inclusion of thin TMD layers in van der Waals heterostructures makes it possible to observe many-particle effects in systems with long carrier lifetime. Due to these factors, TMD-based structures are considered perfect systems for studying high-temperature electron-hole liquid (EHL).

Usually, energy of one electron-hole ($e$-$h$) pair in EHL is $\left|E_\text{EHL}\right|\gtrsim E_x$, and the critical temperature of the gas--liquid phase transition is $T_c\sim0.1\left|E_\text{EHL}\right|$ \cite{Andryushin1977a, Rice1980, JeffrisKeldysh1988, Tikhodeev1985, Sibeldin2016, Sibeldin2017}. Therefore, it can be expected that EHL will be observed in monolayers of TMDs even at room temperatures.

A number of papers were published in 2019 on EHL in monolayers of TMDs. Let's dwell on them in more detail to highlight the current situation.

Note the paper \cite{Bataller2019}, which presents the results of observation of electron-hole plasma (EHP) in the MoS$_2$ monolayer. Although EHL was not observed, the obtained results are important for further research. In particular, it was demonstrated that photoexcitation of a MoS$_2$ monolayer leads to a direct--indirect bandgap transition, which favors the EHL formation.

It was shown in \cite{Younts2019} that an increase in the intensity of continuous optical excitation of a MoS$_2$ monolayer in the below-gap regime initiates the transition of an insulator exciton gas into a metallic EHP.

A high-temperature strongly coupled EHL with $T_c\simeq500$ K in a suspended monolayer MoS$_2$ was investigated in the work \cite{Yu2019}. A first-order gas--liquid phase transition was observed. With the same excitation, EHL differs from EHP by a higher charge carrier density and, hence, by a higher photoluminescence. In the experiment, a sharp step-like increase in photoluminescence was observed when the pump power increased to $\sim3$~kW/cm$^2$. The EHL incompressibility was demonstrated: with a further increase in the pump power, the area occupied by EHL increased in proportion to the number of generated $e$-$h$ pairs, while the EHL density remained constant. Authors of the work \cite{Yu2019} have observed the appearance of a luminescent ring caused by a phonon wind (see, e.g., the review \cite{Tikhodeev1985} and the references cited therein).

EHL was studied at room temperature in ultrathin photocells graphene--a thin (several monolayers) MoTe$_2$ film--graphene \cite{Arp2019}. When the power of the exciting laser reached $P=3$ mW, the $e$-$h$ pair density became so large that the average distance between them $a_{xx}=1\div3$ nm was comparable to the Bohr radius of excitons $a_x=2.3$ nm \cite{Sun2017}. Under these conditions, the exciton annihilation (the transition of an insulator exciton gas into a metallic EHP) was observed. Then, at a higher value of the laser power $P_c=6$ mW, EHP underwent a gas--liquid transition. The formation of a 2D EHL droplet was proved by the appearance of a characteristic dependence for the photocurrent on the exciting laser power and the photocurrent dynamics on a picosecond scale. A similar behavior was observed during the formation of EHL droplets in traditional semiconductors (Si, Ge, GaAs, etc.) at low temperatures \cite{Rice1980, JeffrisKeldysh1988}.

The theoretical work \cite{Rustagi2018} is devoted to the calculation of the EHL phase diagram in a suspended monolayer MoS$_2$. The work was motivated by the experimental results outlined in the paper \cite{Yu2019}. To calculate the EHL properties, the authors of \cite{Rustagi2018} used a potential describing the Coulomb interaction of charge carriers in films of finite thickness (the Keldysh potential \cite{Rytova1967, Chaplik1971, Keldysh1979}), which has proven itself well to explain the significant deviation of the energy of several first exciton levels from the Rydberg series~\cite{Durnev2018}.

According to \cite{Rustagi2018} estimates, the metal--insulator transition in MoS$_2$ occurs in the liquid phase at a density higher than the critical value for the gas--liquid transition. Note that EHP in an ultrathin MoTe$_2$ film is in many respects similar to EHP in the MoS$_2$ monolayer. Therefore, most likely, the metal--insulator transition precedes the gas--liquid transition in both systems.

In our opinion, such a qualitative discrepancy is caused by the unjustified use of the Keldysh potential in the EHL calculations \cite{Rustagi2018}, which leads to a shift of the critical point towards lower densities and temperatures. When calculating the exciton and EHL, the main contributions come from substantially different perturbation theory diagrams, ladder and loop diagrams, respectively.

The influence of the dielectric environment, as well as electron doping (when the densities of electrons and holes are not equal) on the EHL phase diagram (gas--liquid and metal--insulator transitions) were investigated in \cite{Steinhoff2017} by the method of spectral functions. The EHL phase diagram in doped multi-valley semiconductors was considered earlier in the works \cite{Andryushin1990a, Andryushin1991, Andryushin1990b}.

We showed in our previous work \cite{Pekh2020} that the formation of a high-temperature strongly compressed and strongly coupled EHL in monolayer heterostructures based on TMDs is due to the weakening of the Coulomb interaction screening. The Coulomb interaction of charge carriers in monolayer films is determined by the half-sum of permittivities of the environment (for a sample on a substrate, this is a vacuum and a substrate, and for a suspended sample, this is a vacuum). In both cases, the effective permittivity is significantly lower than that of TMD films. As a result, EHL is formed with an anomalously high ground state energy $\left|E_\text{EHL}\right|$, which is fractions of eV, and a high density $n_\text{EHL}$. For example, experimental values for a suspended monolayer MoS$_2$ are $\left|E_\text{EHL}\right|=0.48$ eV and $n_\text{EHL}=4\times10^{13}$ cm$^{-2}$ \cite{Younts2019}.

Based on our previous results, in this paper we investigate the EHL phase diagram in quasi-2D TMD-based heterostructures.

EHL appears through a first-order phase transition, when the density of free and/or bound carriers in excitons reaches a certain temperature-dependent critical value $n_\text{G}(T)$ of the saturated vapor density \cite{Andryushin1977a}. At the critical temperature $T_c$, the critical $e$-$h$ pair density $n_\text{G}(T_c)=n_c$ is reached.

In this work, we investigate the metal--insulator transition (Mott transition), which occurs at the $e$-$h$ pair density $n_{dm}(T)$. With increasing density, this transition usually precedes the gas--liquid transition ($n_{dm}(T)<n_c$) and is in the gas phase \cite{Andryushin1977a, Rice1980}.

We will search for semiconductors in whose films EHL will be characterized by the highest binding energy and, therefore, the highest critical temperature. For this, we will consider model semiconductors with different numbers of electron and hole valleys and unequal effective masses of electrons and holes.

\section{\label{s2}Model foundations}

We describe a model quasi-2D nonequilibrium $e$-$h$ system by the Hamiltonian \cite{Andryushin1976a, Andryushin1976b}
\begin{equation*}
\begin{split}
&\widehat{H}=\sum_{\mathbf{p}sk}^{\nu_e}\frac{1+s}{2}\mathbf{p}^2a^\dag_{\mathbf{p}sk} a_{\mathbf{p}sk}+\sum_{\mathbf{p}sl}^{\nu_h}\frac{1-s}{2}\mathbf{p}^2b^\dag_{\mathbf{p}sl}b_{\mathbf{p}sl}+\\
&+\frac{1}{2}\sum_{\mathbf{p}\mathbf{p}^\prime\mathbf{q}ss^\prime}V(\mathbf{q})\left\{\sum_{kk^\prime}^{\nu_e}a^\dag_{\mathbf{p}sk}a^\dag_{\mathbf{p}^\prime s^\prime k^\prime}a_{\mathbf{p}^\prime+\mathbf{q}s^\prime k^\prime}a_{\mathbf{p}-\mathbf{q}sk}+\right.\\
\end{split}
\end{equation*}
\begin{equation}\label{Ham}
+\sum_{ll^\prime}^{\nu_h}b^\dag_{\mathbf{p}sl}b^\dag_{\mathbf{p}^\prime s^\prime l^\prime}b_{\mathbf{p}^\prime+\mathbf{q}s^\prime l^\prime}b_{\mathbf{p}-\mathbf{q}sl}-
\end{equation}
\begin{equation*}
\left.-2\sum_{kl}^{\nu_e\nu_h}a^\dag_{\mathbf{p}sk}b^\dag_{\mathbf{p}^\prime s^\prime l}b_{\mathbf{p}^\prime+\mathbf{q}s^\prime l}a_{\mathbf{p}-\mathbf{q}sk}\right\}.
\end{equation*}
Here $a_{\mathbf{p}sk}$ ($a^\dag_{\mathbf{p}sk}$) and $b_{\mathbf{p}sl}$ ($b^\dag_{\mathbf{p}sl}$) are Fermi operators of annihilation (creation) of an electron and a hole with quasimomentum $\mathbf{p}$ and spin $s$ in the valleys $k$ and $l$, respectively. The number of electron valleys is $\nu_e$, and the number of hole valleys is $\nu_h$. In general case, $\nu_e\neq\nu_h$.

The normalized area of the layer is set equal to one. We use a system of units with Planck's constant $\hbar=1$ and twice the reduced mass is $2m=1$ ($m^{-1}=m^{-1}_e+m^{-1}_h$). In this work, along with the ratio of the effective masses of the electron and hole $\sigma=m_e/m_h$, we use the parameter $s=(1-\sigma)/(1+\sigma)$.

To take into account the interaction in EHL, we use here the usual 2D Coulomb potential $V(\mathbf{q})=2\pi\widetilde{e}^2/|\mathbf{q}|$, rather than the Keldysh potential \cite{Rustagi2018, Rytova1967, Chaplik1971, Keldysh1979, Pekh2020}
\begin{equation}\label{keldysh_int}
V(\mathbf{q})=\frac{2\pi\widetilde{e}^2}{|\mathbf{q}|(1+r_0|\mathbf{q}|)},
\end{equation}
where the ``screening length'' $r_0=d/2\delta$, $d$ is the film thickness, $\delta=\varepsilon_\text{eff}/\varepsilon$, $\varepsilon$ is the permittivity of the film material, $\varepsilon_\text{eff}=(\varepsilon_1+\varepsilon_2)/2$ is the effective permittivity of the media surrounding the film (e.g., $\varepsilon_1=1$ for vacuum and $\varepsilon_2=3.9$ for the SiO$_2$ substrate). As applied to TMD monolayer systems, the quantity $r_0$ is an adjustable parameter in the exciton calculations \cite{Durnev2018}.

In the previous work \cite{Pekh2020}, we showed that the use of the Keldysh potential for calculating EHL is not justified. With the help of one adjustable parameter, it is impossible to agree both calculated characteristics with the experimental ones: the binding energy of EHL and its equilibrium density.

We also put $\widetilde{e}^2=1$, when defining the system of units. Then, the binding energy and the radius of the 2D exciton are automatically equal to unity: $E_x=2m\widetilde{e}^4/\hbar^2=1$ and $a_x=\hbar^2/2m\widetilde{e}^2=1$. Below, it is assumed that the energy and temperature are measured in $E_x$, and the $e$-$h$ pair density are measured in $a^{-2}_x$.

In the next two sections, we will focus on the calculation of the ground state energy of EHL and its thermodynamics in the TMD monolayer, using, among other things, the previously obtained results for quasi-2D systems \cite{Klyuchnik1978, Andryushin1979, Silin1988}.

\section{\label{s3}Ground state energy}

We are considering a semiconductor in which the number of electron and hole valleys is not necessarily equal. In what follows, without loss of generality, we will assume that $\nu_e\geq\nu_h$ (i.e., we call charge carriers with the largest number of valleys electrons). The momenta $q$ will be measured in units of $q_F=q^e_F\leq q^h_F$, where $q^e_F$ and $q^h_F$ are the Fermi momenta of electrons and holes, respectively, $q^{e,h}_F=\sqrt{2\pi n/\nu_{e,h}}$ ($n$ is the 2D charge carrier density). Frequency $\omega$ will be measured in units of $q^2_F/2m$. We introduce the dimensionless distance between the particles $r_s=\sqrt{\nu_e/\pi n}$.

Hereinafter, we assume that there is spin degeneracy for both electrons and holes.

In some cases, the spin-orbit splitting of the valence band in TMD is very large (e.g., in MoS$_2$ it is equal to 148 meV \cite{Rustagi2018}), then $\nu_h/2$ should be used instead of $\nu_h$.

We also recall that under intense photoexcitation of the MoTe$_2$ film, holes flow down to the $\Gamma$ point and $\nu_h=1$, while the number of electron valleys does not change and $\nu_e=2$ \cite{Arp2019}.

The ground state energy of 2D EHL per one $e$-$h$ pair is
\begin{equation}\label{E_gs}
E_\text{gs}=E_\text{kin}+E_\text{exch}+E_\text{corr}.
\end{equation}
The first term is the kinetic energy ($\varkappa=\nu_e/\nu_h$ and $\varkappa\geq1$)
\begin{equation}\label{E_kin}
E_\text{kin}=\frac{1+\varkappa\sigma}{1+\sigma}r^{-2}_s.
\end{equation}
The second term is the exchange energy
\begin{equation}\label{E_exch}
E_\text{exch}=-\frac{4\sqrt{2}}{3\pi}\left(1+\sqrt{\varkappa}\right)r^{-1}_s.
\end{equation}
The third term is the correlation energy, which we represent as an integral over the momentum transfer \cite{Pekh2020, Combescot1972, Andryushin1976a, Andryushin1976b, Andryushin1977b}
\begin{equation}\label{corr_e}
E_\text{corr}=\int\limits_0^\infty I(q)dq.
\end{equation}

\begin{figure}[b!]
\begin{center}
\includegraphics[width=0.17\textwidth]{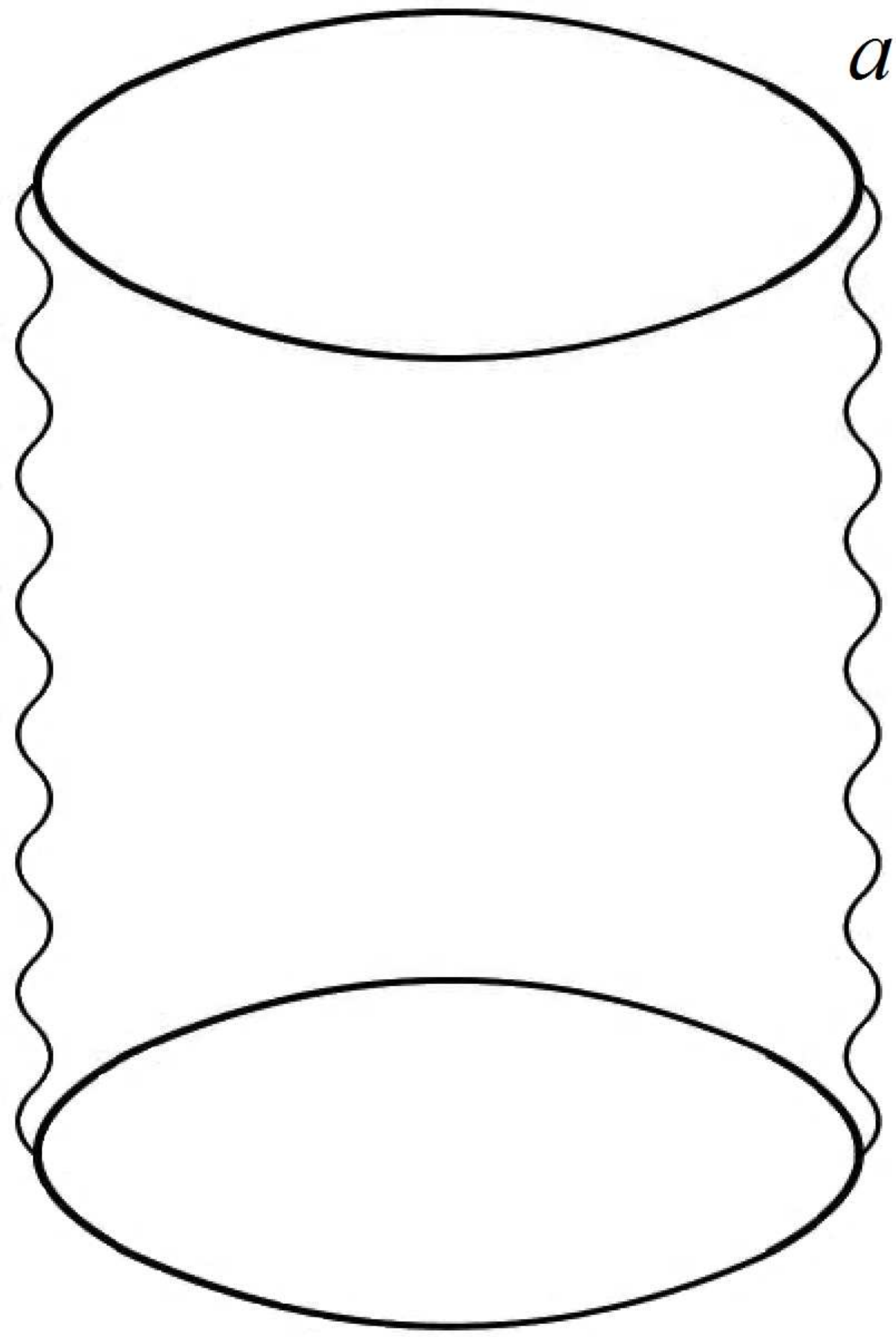}\hspace{0.06\textwidth}\includegraphics[width=0.17\textwidth]{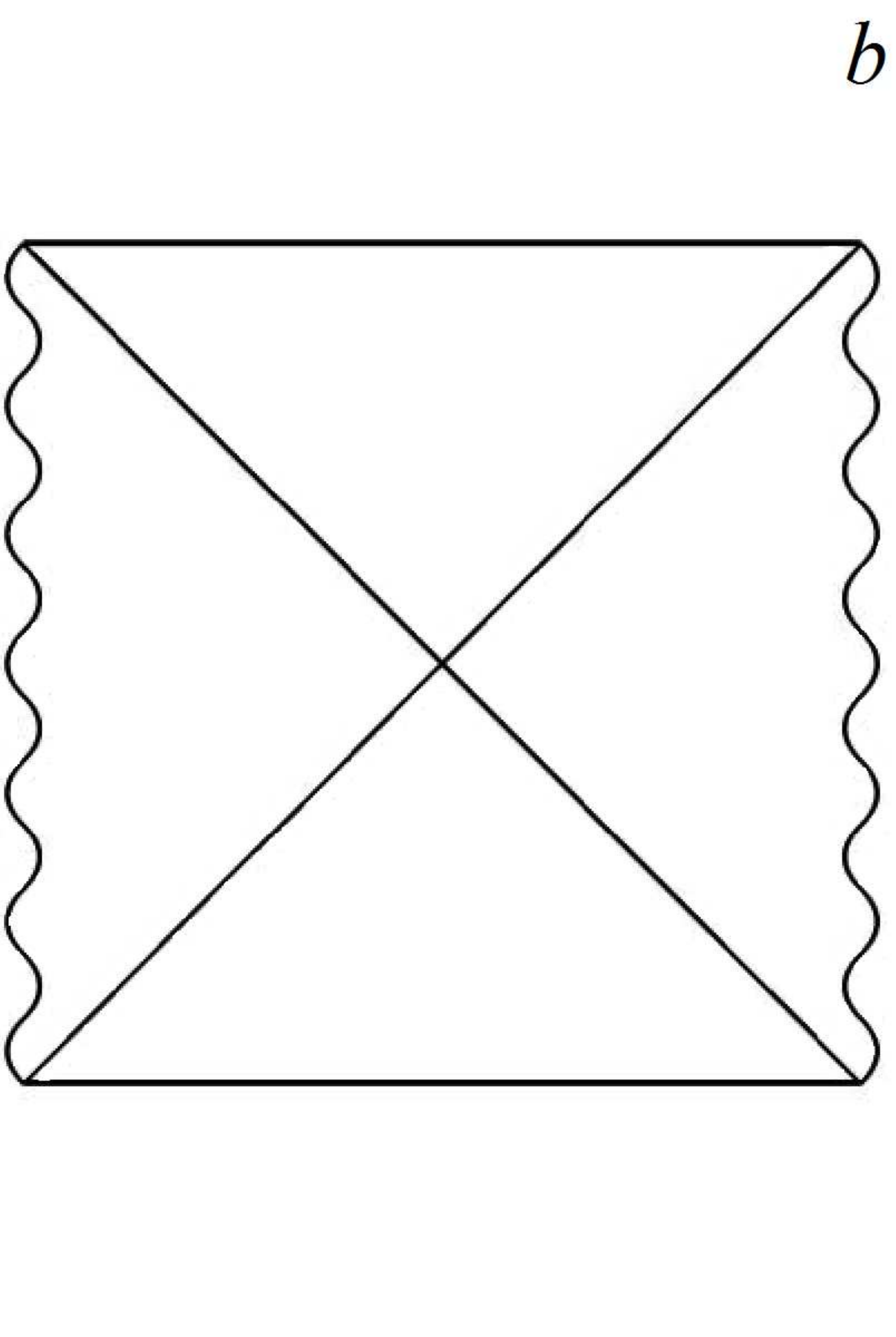}
\caption{\label{f1} Diagrams of the second order in the interaction, defining the function $I(q)$ for $q\gg1$: $a$) direct loop diagram; $b$) exchange diagram.}
\end{center}
\end{figure}

For $q\ll1$, we use the random phase approximation (RPA) \cite{Combescot1972, Andryushin1976a, Andryushin1976b, Andryushin1977b}. The function $I(q)$ for $q\gg1$ is determined by the sum of diagrams of the second order in the interaction (Fig.~\ref{f1}). In the intermediate region, the function $I(q)$ is approximated by a segment of the tangent \cite{Pekh2020, Andryushin1976a, Andryushin1976b}.

For arbitrary values of $\sigma$ and $\varkappa$, the expansion of the function $I(q)$ for small $q$ turns out to be rather cumbersome. For the case of equal masses of an electron and a hole ($\sigma=1$) and an equal number of electron and hole valleys ($\varkappa=1$), it is given in our previous work \cite{Pekh2020}.

The terms in $I(q)$ at $q\ll1$ with half-integer powers of $q$ are the contribution from the 2D plasmon
\begin{equation}\label{plasmon}
\omega(q)=2^{1/4}(\nu_er_sq)^{1/2}+\frac{3}{2^{5/4}(\nu_er_s)^{1/2}}\left(\frac{1}{\eta^3_e}+\frac{\varkappa}{\eta^3_h}\right)q^{3/2}.
\end{equation}
Here, $\eta_e=m_e/m=1+\sigma$ and $\eta_h=m_h/m=1+\sigma^{-1}$.

It is interesting to note that, in contrast to the three-dimensional (3D) case, in the 2D case there is also a branch of damped excitations
\begin{equation}\label{plasmon1}
\omega_1(q)=\frac{2}{\eta_e}q-\frac{1}{\eta_e}q^2+\frac{1}{\eta_e}\left(1+\frac{1}{\varkappa\sigma}\right)^2q^3,
\end{equation}

When the inequalities
\begin{equation}\label{condit_plasm}
\begin{split}
1+\varkappa\sigma&<\sigma\sqrt{\varkappa(1+2\varkappa\sigma)}\\
1+\varkappa\sigma&>\frac{\sigma^2\varkappa^{3/2}}{\sqrt{\sigma^2\varkappa-1}}
\end{split}
\end{equation}
are satisfied, two branches of damped oscillations are also possible
\begin{equation}\label{plasmon2}
\begin{split}
&\omega_2(q)=\frac{2(1+\varkappa\sigma)}{\eta_e\sqrt{1+2\varkappa\sigma}}q-\frac{2\sqrt{2}\varkappa^3\sigma^3q^2}{\nu_e\eta_e^2r_s(1+2\varkappa\sigma)^{3/2}}\\
&-\frac{2(1+\varkappa\sigma)}{\eta_e\sqrt{1+2\varkappa\sigma}}\left[\frac{3\varkappa^6\sigma^6}{\nu^2_e\eta_e^2r^2_s(1+2\varkappa\sigma)^3}+\frac{1}{8}\right]q^3,
\end{split}
\end{equation}
\begin{equation}\label{plasmon3}
\begin{split}
&\omega_3(q)=\frac{2\sqrt{\varkappa}}{\eta_h}q-\frac{1}{\eta_h}q^2\\
&+\frac{1}{\eta_h\sqrt{\varkappa}}\left(1+\varkappa\sigma-\frac{\sigma^2\varkappa^{3/2}}{\sqrt{\sigma^2\varkappa-1}}\right)^2q^3.
\end{split}
\end{equation}
We assumed that the most frequent case $\sigma\sqrt{\varkappa}>1$ is realized (for many TMDs $\sigma=0.73\div1.12$ and for $\nu_e\neq\nu_h$ the relation $\varkappa\geq2$).

\begin{figure}[b!]
\begin{center}
\includegraphics[width=0.5\textwidth]{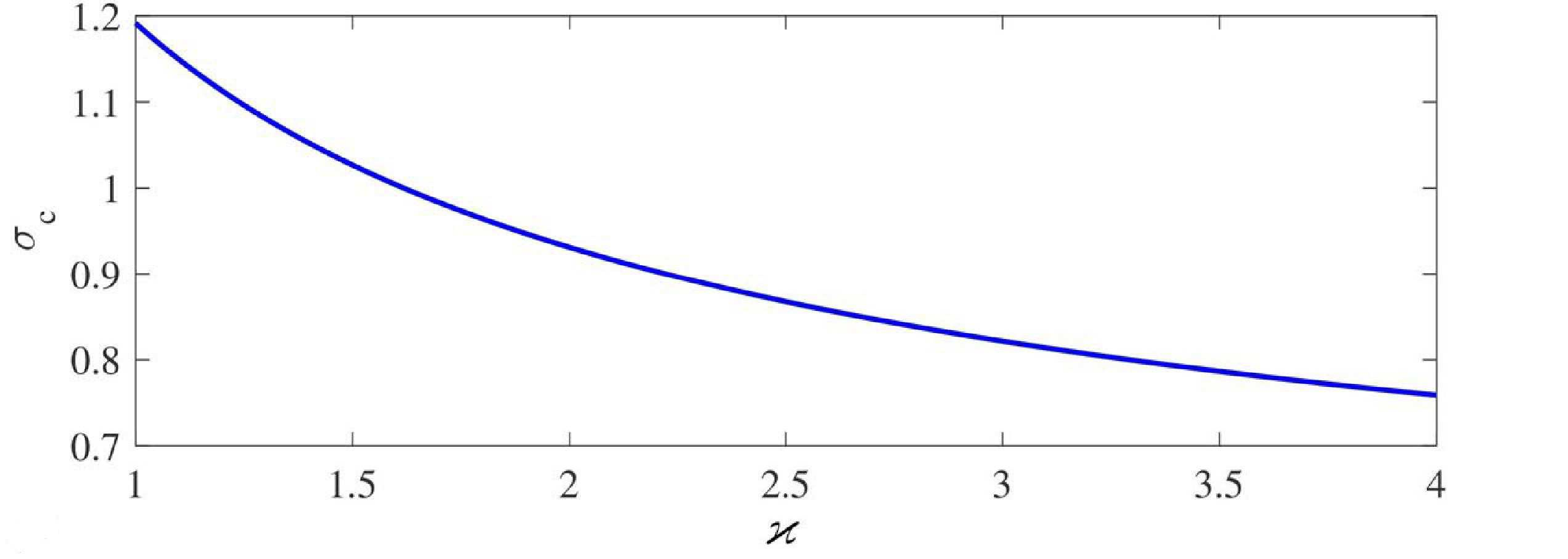}
\end{center}
\caption{\label{f2} Numerical calculation $\sigma_c(\varkappa)$.}
\end{figure}

The inequalities \eqref{condit_plasm} are satisfied for $\sigma>\sigma_c(\varkappa)$,
\begin{equation}\label{sigma_c}
\sigma_c(\varkappa)\approx\frac{1.032\varkappa^{3/4}}{\varkappa-0.136}.
\end{equation}
In particular, $\sigma_c(2)=0.931$. The graph of the function $\sigma_c(\varkappa)$ is shown in Fig.~\ref{f2}.

Apparently, the presence of additional branches \eqref{plasmon1}, \eqref{plasmon2}, \eqref{plasmon3}, which are absent in the 3D case, provides 2D EHL binding.

The EHL equilibrium energy $E_\text{EHL}$ is the minimum of the ground state energy $E_\text{gs}$
\begin{equation*}
\left.\frac{\partial E_\text{gs}}{\partial n}\right|_{n_\text{EHL}}=0,~E_\text{EHL}=E_\text{gs}(n_\text{EHL}),
\end{equation*}
where $n_\text{EHL}$ is the equilibrium EHL density.

The equation for $n_\text{EHL}$
\begin{equation*}
\frac{1+\varkappa\sigma}{1+\sigma}\frac{\pi}{\nu_e}-\frac{2\sqrt{2}(1+\sqrt{\varkappa})}{3\sqrt{\pi\nu_en_\text{EHL}}}+\left.\frac{\partial E_\text{corr}}{\partial n}\right|_{n_\text{EHL}}=0.
\end{equation*}
is very cumbersome and for the sake of brevity we do not present it here. In the case of an equal number of electron and hole valleys, to determine $n_\text{EHL}$ and $E_\text{EHL}$, one can use, respectively, formulas (21) and (22) of our work \cite{Pekh2020}.

\section{\label{s4}Monolayer thermodynamics}

In keeping with the analogy of nonequilibrium charge carriers in semiconductors with electrons and ions in crystals, L.~V.~Keldysh suggested \cite{Keldysh1968} that with a decrease in temperature or an increase in the density of excitons or exciton molecules (biexcitons), this gas will be liquefied into EHL and the system will undergo a first-order gas--liquid phase transition. Since the late 1960s, the EHL formation has been observed in many semiconductors (see, e.g., \cite{Rice1980, JeffrisKeldysh1988, Tikhodeev1985, Sibeldin2016, Sibeldin2017} and references therein).

The gas--liquid transition corresponds to a bell-shaped curve on the $(n,\,T)$ plane with a critical point $(n_c,\,T_c)$ in the top. This curve passes through the points $n=n_0$ and $n=n_\text{EHL}$ at $T=0$. The value of $n_0$ is determined by the equality $|E_\text{EHL}|=E_x+\frac{1}{2}E_D$, where $E_D$ is the biexciton dissociation energy \cite{JeffrisKeldysh1988}. The region to the left of the $n_\text{G}$ branch emerging from the point $(n_0,\,0)$ corresponds to a (bi)exciton insulator gas; to the right of the branch $n_\text{L}$, emerging from the point $(n_\text{EHL},\,0)$, there is an EHL region. Under the ``bell'' there is an unstable region in which separation into EHL droplets with a density $n_\text{L}$ and a saturated vapor of excitons or $e$-$h$ pairs with a density $n_\text{G}$ occurs.

The phase diagram also contains a second bell-shaped curve, which corresponds to the exciton--plasma transition (metal--insulator transition). The construction of this curve for arbitrary densities and temperatures is difficult and we restrict ourselves to estimates in the most interesting areas. In the classical region ($T\gg E_F$, $E_F$ is the Fermi energy of $e$-$h$ pairs), this curve is determined by the reaction of exciton formation from a free electron and a free hole. As the density increases at a given temperature, electrons and holes are bound to form excitons. In the quantum region ($T\ll E_F$), an increase in the density of excitons leads to their dissociation (a Mott transition occurs). The possibility of a metal--insulator transition in the liquid phase was discussed in the works \cite{Bisti1978, Lobaev1984}.

In the region $T\sim E_x$, excitons undergo thermal dissociation. Therefore, the transition of excitons to EHP is significantly smeared out and the exciton spectra at approaching $n_c$ have a diffuse character \cite{Kukushkin1983}.

The most promising for the observation of the exciton--plasma transition are semiconductors, in which the EHL stability decreases and $T_c$ decreases due to the lifting of the band degeneracy. Due to this, in such semiconductors at sufficiently low temperatures, it is possible to realize excitonic gas densities close to $n_c$. For example, it was experimentally found in uniaxially deformed Ge that the metal--insulator transition occurs at $r_s=2\pm0.1$. This is in good agreement with the estimate $r_s\approx1.8$ obtained using dielectric screening of excitons \cite{Bisti1978}.

The three-component system (EHL, free excitons, and free charge carriers) was considered in the works \cite{Keldysh1974, Sanina1985}.

Calculations of the thermodynamic properties of EHL in quasi-2D systems were carried out earlier in works \cite{Klyuchnik1978, Andryushin1979, Silin1988}. Materials such as TMD monolayers and graphene are 2D crystals. The thermodynamics of $e$-$h$ systems is equally applicable to both quasi-2D heterostructures and 2D crystals. However, for the latter, there is a doubt about their stability with respect to vibrations of atoms across the plane of the crystal \cite{Lozovik2008}. In addition, in the process of intense photoexcitation, local overheating of the sample can occur. This can lead to a local deformation of the crystal lattice and to melting of 2D crystal \cite{Berezinskii1970, Berezinskii1971, KostelitzThouless1973, NelsonHalperin1978, NelsonHalperin1979}. The experimentally observed intense photoluminescence allows us to conclude that there are almost no defects in TMD monolayers. Nonradiative recombination of charge carriers at defects would lead to luminescence quenching. We believe that the melting temperature significantly exceeds the critical temperature of the gas--liquid transition $T_c$, and intense photoexcitation of 2D crystal can only lead to the lifting of the valley degeneracy, which was observed, for example, in an ultrathin film MoTe$_2$ \cite{Arp2019}.

We represent the chemical potential of an $e$-$h$ pair ensemble in a TMD monolayer in the form \cite{Andryushin1979, Silin1988}
\begin{equation*}
\mu(n,\,T)=T\ln\left[\left(e^{\frac{2\pi n}{(1+\sigma)\nu_eT}}-1\right)\left(e^{\frac{2\pi\sigma n}{(1+\sigma)\nu_hT}}-1\right)\right]
\end{equation*}
\begin{equation}\label{mu_general}
-\frac{2\sqrt{2}}{\sqrt{\pi}}\left(\frac{1}{\sqrt{\nu_e}}+\frac{1}{\sqrt{\nu_h}}\right)\sqrt{n}+\frac{\partial}{\partial n}\left(nE_\text{corr}\right).
\end{equation}

The right-hand side of \eqref{mu_general} is the chemical potential of noninteracting particles $\mu_\text{kin}$ (first term) together with exchange $\mu_\text{exch}$ and correlation $\mu_\text{corr}$ contributions (second and third terms, respectively). We have neglected in \eqref{mu_general} the dependence of $\mu_\text{exch}$ and $\mu_\text{corr}$ on temperature, since the temperature corrections to $\mu_\text{exch}$ and $\mu_\text{corr}$ cancel each other at $T\ll E_F$. This is in agreement with the previously obtained results for the 3D case \cite{JeffrisKeldysh1988, Thomas1974, Rice1974}.

The critical point $(n_c,\,T_c)$ of the gas--liquid transition is determined by two equations
\begin{equation}\label{crit_point}
\left.\frac{\partial\mu}{\partial n}\right|_{\substack{n=n_c\\ T=T_c}}=\left.\frac{\partial^2\mu}{\partial n^2}\right|_{\substack{n=n_c\\ T=T_c}}=0.
\end{equation}

To test our methods for studying the EHL phase diagram in TMD monolayers, we calculated the similarity relation typical for the gas--liquid transition in general (see the note on p. 262 of the book \cite {Landau5}) $T_c\simeq0.1|E_\text{EHL}|$ \cite{Andryushin1977a}, as well as the ratio of the critical density to the equilibrium one $n_c/n_\text{EHL}\approx0.2$ \cite{JeffrisKeldysh1988}.

Table~\ref{t1} shows the results of the numerical solution of the equations \eqref{crit_point} for $\sigma=1$. Note that for a relatively small number of valleys, the ratio $|E_\text{EHL}|/T_c$ is almost independent of the number of valleys. The maximum of this ratio is attained for the system $\nu_e=\nu_h=4$, which is possible for alloys of the form Mo$_x$W$_{1-x}$S$_2$ with equivalent energy valleys MoS$_2$ and WS$_2$.

\begin{table}[b!]
\caption{\label{t1} Critical point of gas--liquid transition and 2D EHL parameters in TMD monolayers.}
\begin{ruledtabular}
\begin{tabular}{cccccc}
$(\nu_e,~\nu_h)$ & (1,~1) & (2,~1) & (2,~2) & (4,~1) & (4,~4)\\
\hline
$n_c$ & 0.035 & 0.054 & 0.089 & 0.075 & 0.242 \\
$T_c$ & 0.136 & 0.164 & 0.196 & 0.198 & 0.294 \\
$n_\text{EHL}$ & 0.164 & 0.264 & 0.450 & 0.394 & 1.302 \\
$|E_\text{EHL}|$ & 1.090 & 1.340 & 1.661 & 1.606 & 2.545 \\
$n_\text{EHL}/n_c$ & 4.686 & 4.889 & 5.056 & 5.253 & 5.379 \\
$|E_\text{EHL}|/T_c$ & 8.015 & 8.171 & 8.111 & 8.070 & 8.718 \\
\end{tabular}
\end{ruledtabular}
\end{table}

\begin{figure}[t!]
\begin{center}
\includegraphics[width=0.25\textwidth]{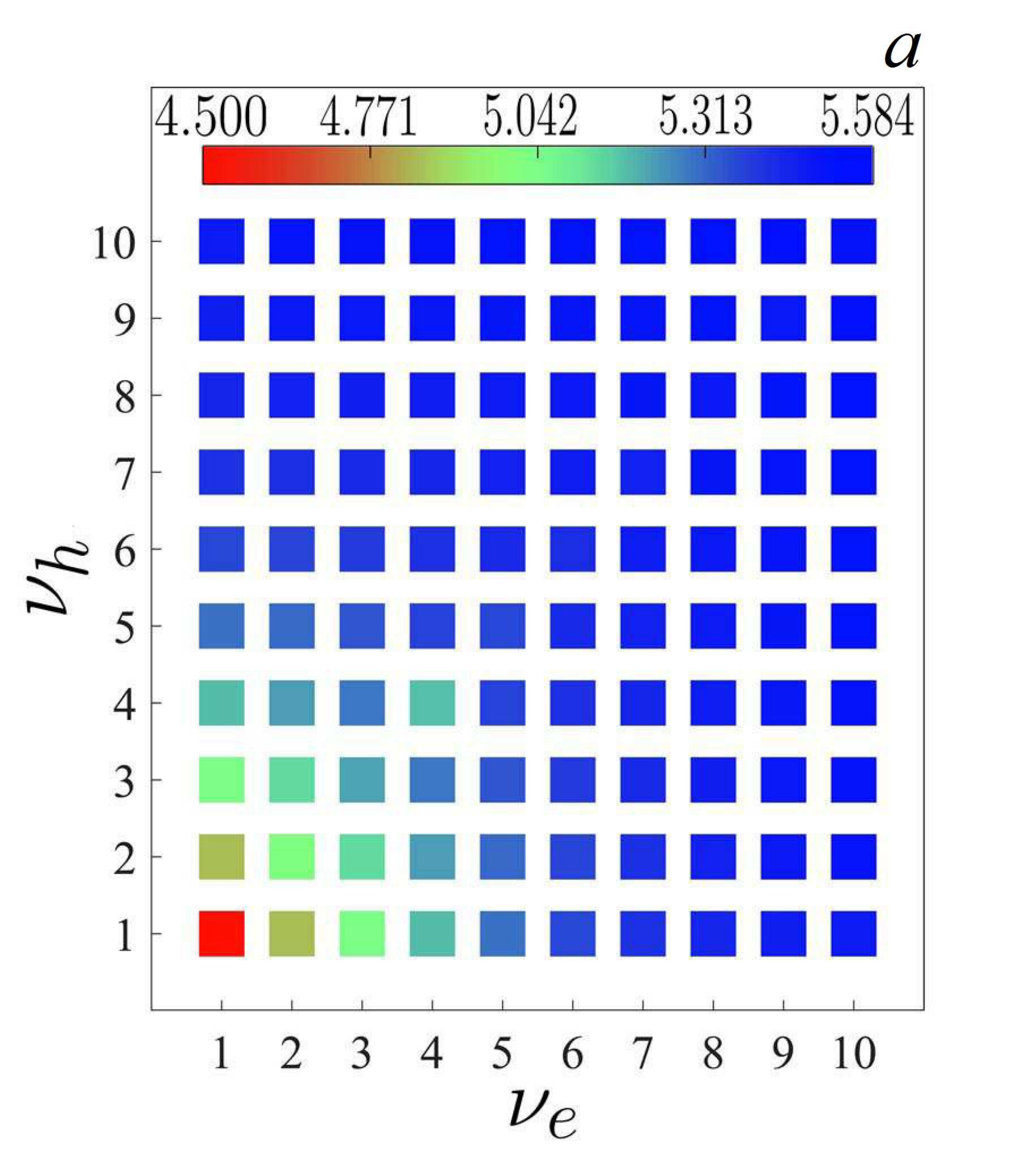}\includegraphics[width=0.25\textwidth]{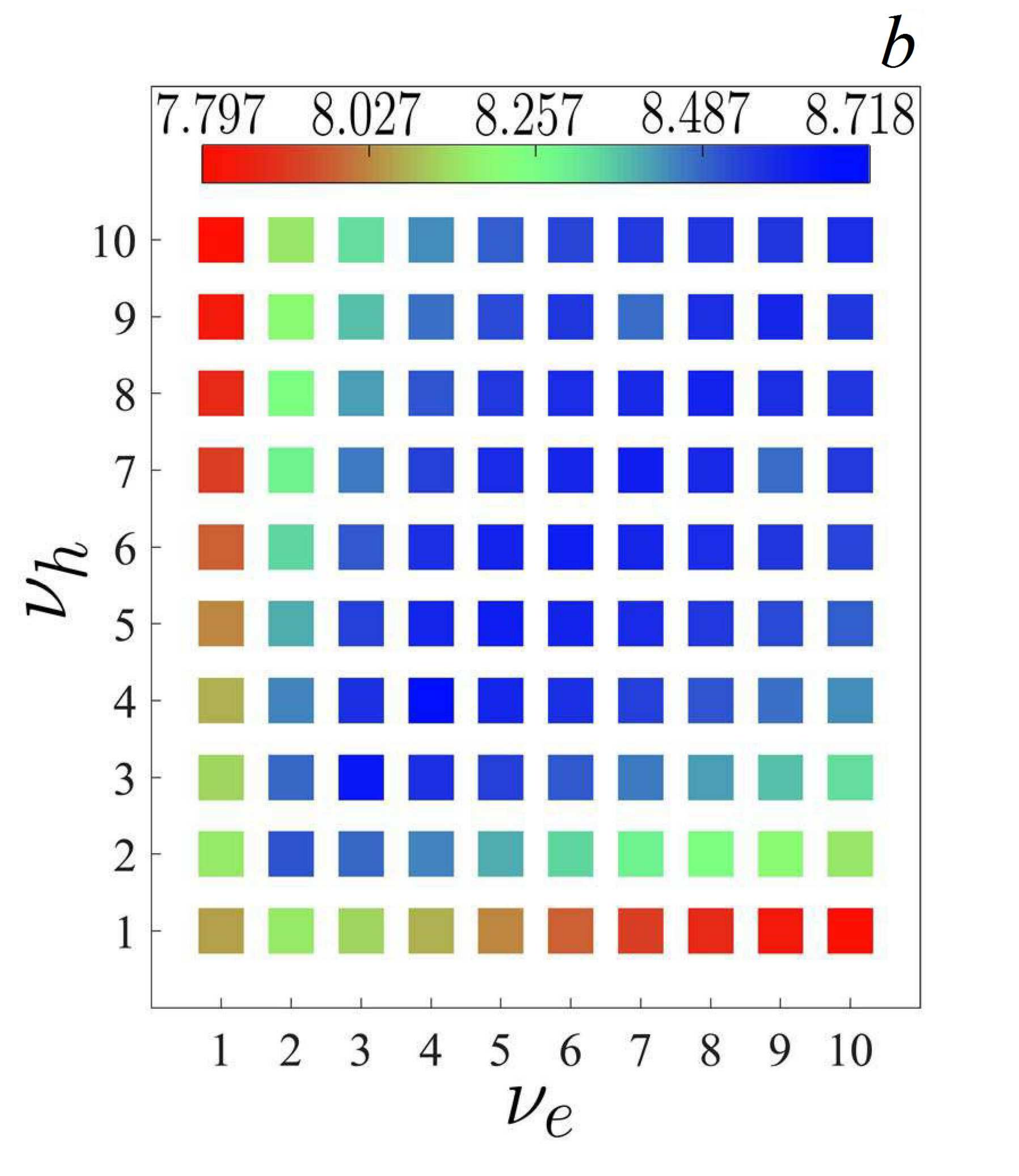}
\includegraphics[width=0.25\textwidth]{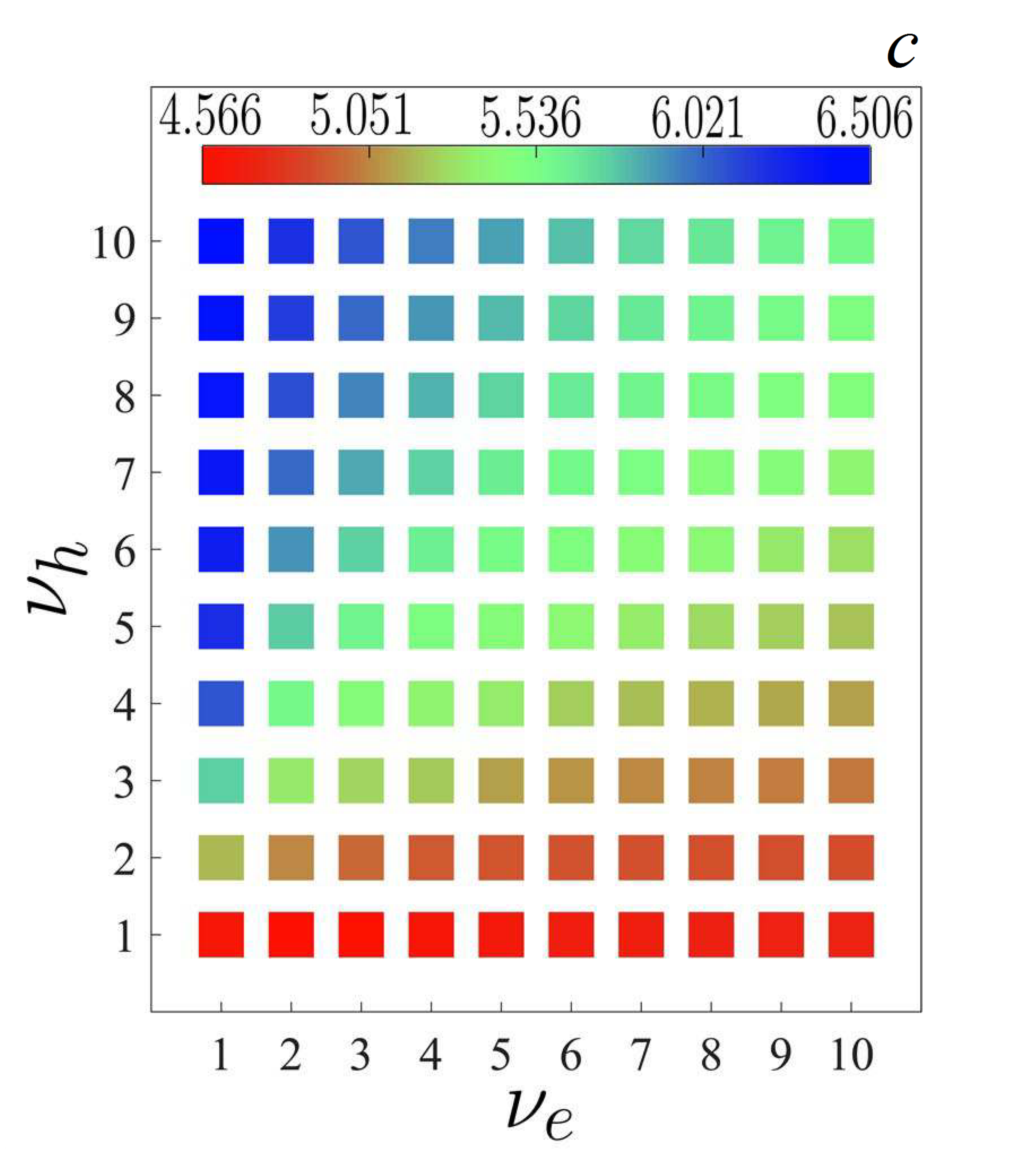}\includegraphics[width=0.25\textwidth]{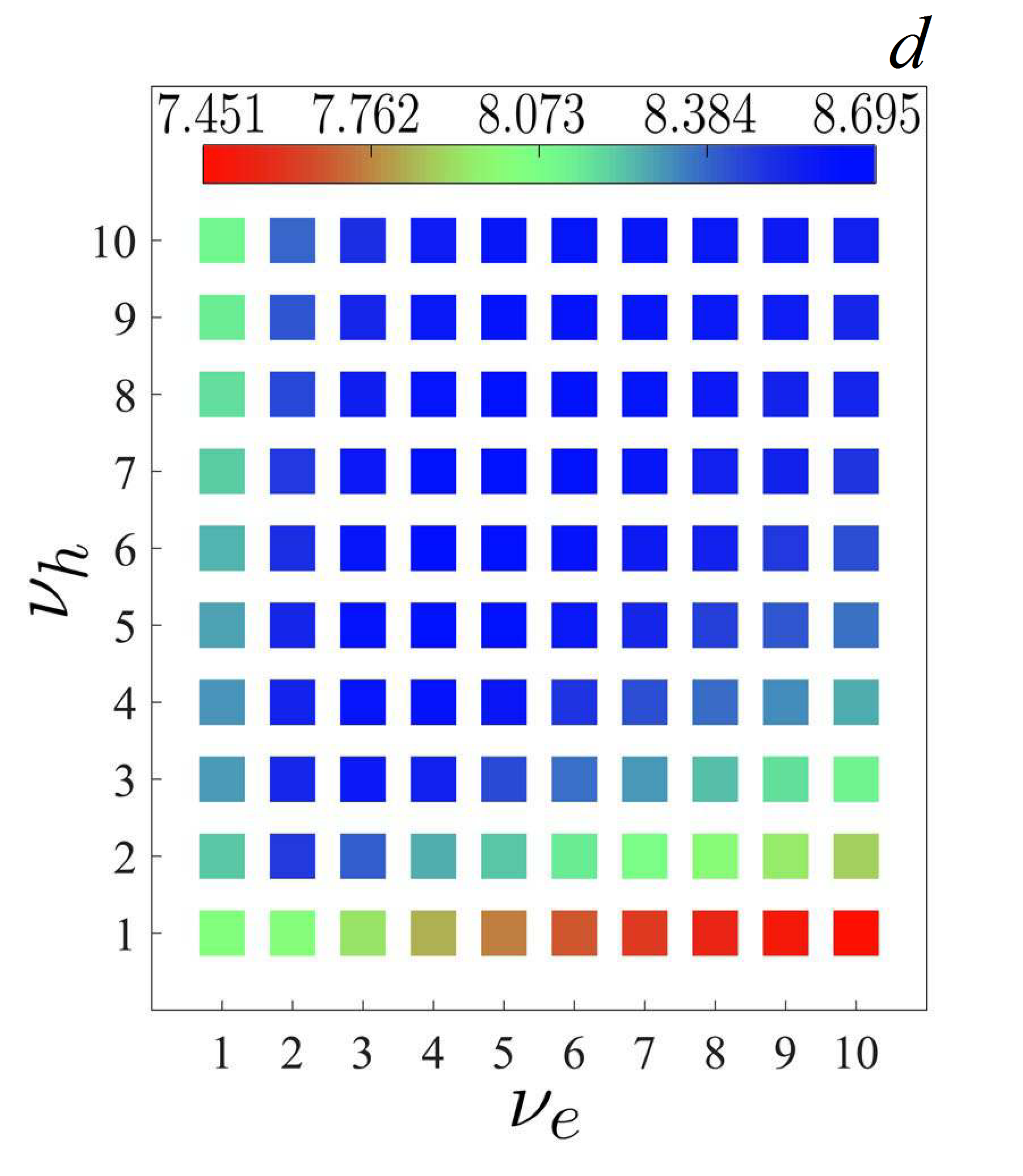}
\end{center}
\caption{\label{f3} Dependence of the ratios $n_\text{EHL}/n_c$ and $|E_\text{EHL}|/T_c$ on the number of electronic $\nu_e$ and hole $\nu_h$ valleys: $a$) $n_\text{EHL}/n_c$ at $\sigma=1$; $b$) $|E_\text{EHL}|/T_c$ at $\sigma=1$; $c$) $n_\text{EHL}/n_c$ at $\sigma=0.8$; $d$)~$|E_\text{EHL}|/T_c$ at $\sigma=0.8$.}
\end{figure}

The results of the numerical calculation of the ratios $n_\text{EHL}/n_c$ and $|E_\text{EHL}|/T_c$ for a different number of valleys are shown in Fig.~\ref{f3}.

For $\sigma=1$, the area $\nu_e\approx\nu_h\gg1$ (the upper right corner on the upper panels of Fig.~\ref{f3}) corresponds to close to maximum $n_\text{EHL}/n_c$ and $|E_\text{EHL}|/T_c$.

For $\nu_e=1,~\nu_h\gtrsim5$ or for $\nu_h=1,~\nu_e\gtrsim5$ $n_\text{EHL}/n_c$ is close to the maximum, and $|E_\text{EHL}|/T_c$ tends to the minimum. This is due to the faster growth of $T_c$ compared to $|E_\text{EHL}|$ in these areas.

At $\sigma\neq1$ (here $\sigma=0.8$ is taken, see the bottom left panel of Fig.~\ref{f3}) $n_\text{EHL}/n_c$ changes noticeably: the minimum moves to the region with large $\nu_e$ and small $\nu_h$; the~maxima remain in the region of large $\nu_h$ and small~$\nu_e$; for $\nu_e\approx\nu_h\gg1$, the values are intermediate. It is interesting to note that there is no such restructuring for the more fundamental relation $|E_\text{EHL}|/T_c$, and changes occur mainly in the region $\nu_e=1,~\nu_h\gtrsim5$ (see lower right panel Fig.~\ref{f3}).

In the limit of a large number of valleys ($\nu_e=\nu_h=\nu\rightarrow\infty$), the correlation energy in the expression \eqref{mu_general} is a power density function
\begin{equation}\label{Ecorr_An_gamma}
E_\text{corr}=-A(\sigma)n^\gamma.
\end{equation}
For 2D systems, the exponent is $\gamma=\hspace{0.05cm}^1\hspace{-0.08cm}/_3$. The $A(\sigma)$ function was defined by us in the previous work \cite{Pekh2020}. In particular, $A(1)\approx4.774$.

In the limit $\nu\rightarrow\infty$, the exchange contribution to both the energy and the chemical potential can be neglected. In this case, the pair of equations \eqref{crit_point} is reduced \cite{Andryushin1977a} to the equation
\begin{equation}\label{eq_zc}
\begin{split}
&z_c=(1-\gamma)\\
&\times\frac{\left(\cosh z_c-\cosh(sz_c)\right)\left(e^{z_c}-\cosh(sz_c)-s\sinh(sz_c)\right)}{(1+s^2)\left(\cosh z_c\cosh(sz_c)-1\right)-2s\sinh z_c\sinh(sz_c)}
\end{split}
\end{equation}
with variable $z_c=\frac{\pi n_c}{\nu T_c}$. Here, $s=\frac{1-\sigma}{1+\sigma}$.

The critical parameters of the EHL are expressed in terms of $z_c$ by the formulas
\begin{equation}
n_c=\left[\frac{\gamma(1+\gamma)A\nu}{\pi}\frac{\cosh z_c-\cosh(sz_c)}{e^{z_c}-\cosh(sz_c)-s\sinh(sz_c)}\right]^\frac{1}{1-\gamma},
\end{equation}
\begin{equation}
T_c=\frac{\pi\nu^{\frac{\gamma}{1-\gamma}}}{z_c}\left[\frac{\gamma(1+\gamma)A}{\pi}\frac{\cosh z_c-\cosh(sz_c)}{e^{z_c}-\cosh(sz_c)-s\sinh(sz_c)}\right]^\frac{1}{1-\gamma},
\end{equation}
\begin{equation}
\begin{split}
\mu_c=&-T_c\left\{\left(\frac{1}{\gamma}-1+\frac{1}{\gamma}\frac{\sinh z_c-s\sinh(sz_c)}{\cosh z_c-\cosh(sz_c)}\right)z_c\right.\\
&\left.-\ln2-\ln\left(\cosh z_c-\cosh(sz_c)\right)\right\}.
\end{split}
\end{equation}

\begin{figure}[t!]
\begin{center}
\includegraphics[width=0.5\textwidth]{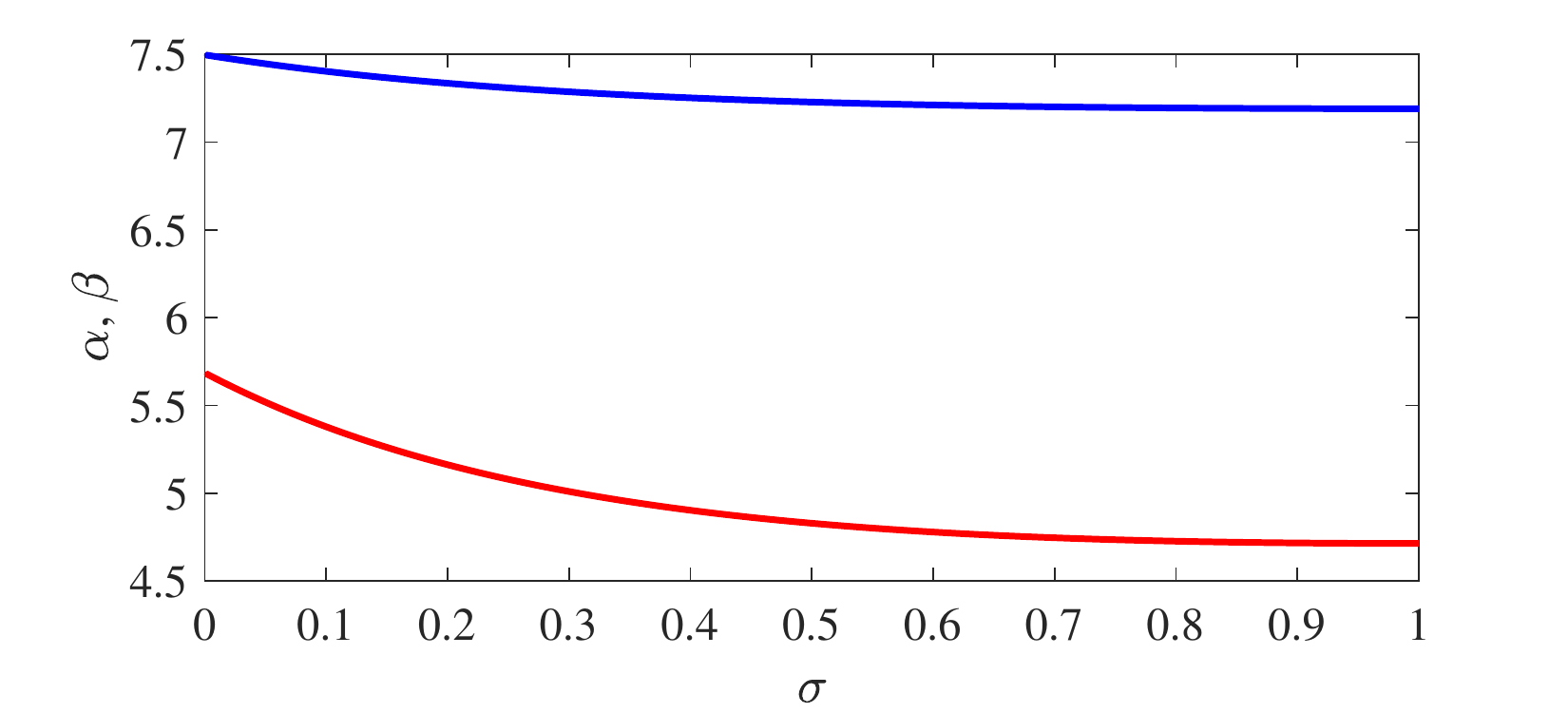}
\end{center}
\caption{\label{f4}  Numerical calculation of the dependence on $\sigma$ of the relations $\alpha=n_\text{EHL}/n_c$ (red curve) and $\beta=|E_\text{EHL}|/T_c$ (blue curve) at $\nu\rightarrow\infty$.}
\end{figure}

Note that in the limit $\nu\rightarrow\infty$ the relations $n_\text{EHL}/n_c$ and $|E_\text{EHL}|/T_c$ do not depend on $A$
\begin{equation}\label{n/nc}
\frac{n_\text{EHL}}{n_c}=\left[\frac{e^{z_c}-\cosh(sz_c)-s\sinh(sz_c)}{(1+\gamma)\left(\cosh z_c-\cosh(sz_c)\right)}\right]^\frac{1}{1-\gamma},
\end{equation}
\begin{equation}\label{E/Tc}
\frac{|E_\text{EHL}|}{T_c}=\frac{1-\gamma}{\gamma}z_c\left[\frac{e^{z_c}-\cosh(sz_c)-s\sinh(sz_c)}{(1+\gamma)\left(\cosh z_c-\cosh(sz_c)\right)}\right]^\frac{1}{1-\gamma}
\end{equation}
and weakly depend on $\sigma$ (see Fig.~\ref{f4}).

In the particular case $\sigma=1$, the equation \eqref{eq_zc} is greatly simplified:
\begin{equation}\label{eq_zc_sigma1}
z_c=(1-\gamma)\left(e^{z_c}-1\right).
\end{equation}
For $\gamma=\hspace{0.05cm}^1\hspace{-0.08cm}/_3$ it has the solution $z_c=0.7627$, while $n_\text{EHL}/n_c=4.7133$ and $|E_\text{EHL}|/T_c=7.1896$,
\begin{equation}
n_c=\left[\frac{\gamma(1+\gamma)A\nu}{2\pi(1-\gamma)}z_ce^{-z_c}\right]^\frac{1}{1-\gamma}\approx0.077\nu^{3/2},
\end{equation}
\begin{equation}
T_c=\left(\frac{\nu z_c}{\pi}\right)^\frac{\gamma}{1-\gamma}\left[\frac{\gamma(1+\gamma)A}{2(1-\gamma)}e^{-z_c}\right]^\frac{1}{1-\gamma}
\approx0.315\nu^{1/2},
\end{equation}
\begin{equation}
\mu_c=-2\left\{\frac{1-\gamma}{\gamma}+\frac{z_c}{\gamma}-\ln\frac{z_c}{1-\gamma}\right\}T_c\approx-8.307T_c.
\end{equation}

The function $z_c(\sigma)$ is obtained with good accuracy by expanding the right-hand side of \eqref{eq_zc} in a series in $z_c$ up to terms of the third order:
\begin{equation}\label{approx_zc}
z_c(\sigma)\approx\frac{3}{2(1+s^2)}\left[\sqrt{1+\frac{8\gamma}{3(1-\gamma)}(1+s^2)}-1\right],
\end{equation}
then
\begin{equation}
n_c\approx\left[\frac{\gamma(1-\gamma^2)A\nu}{2\pi}\left(1+\frac{1+s^2}{12}z^2_c\right)\right]^\frac{1}{1-\gamma},
\end{equation}
\begin{equation}
T_c\approx\frac{\pi\nu^{\frac{\gamma}{1-\gamma}}}{z_c}\left[\frac{\gamma(1-\gamma^2)A}{2\pi}\left(1+\frac{1+s^2}{12}z^2_c\right)\right]^\frac{1}{1-\gamma},
\end{equation}
\begin{equation}
\begin{split}
\mu_c\approx&-T_c\left\{\frac{2}{\gamma(1-\gamma)}\left(1-\frac{1+s^2}{12}z^2_c\right)\right.\\
&\left.-z_c-\ln\left[(1-s^2)z^2_c\left(1+\frac{1+s^2}{12}z^2_c\right)\right]\right\}.
\end{split}
\end{equation}

To obtain the temperature dependence of the $e$-$h$ pair density in the gas $n_\text{G}(T)$ and liquid $n_\text{L}(T)$ phases, we use the Maxwell rule
\begin{equation}\label{Maxwell_rule}
\int\limits_{n_\text{G}}^{n_\text{L}}\mu(n,\,T)dn=\mu(T)(n_\text{L}-n_\text{G}),
\end{equation}
where $\mu(T)=\mu(n_\text{G},~T)=\mu(n_\text{L},~T)$.

\begin{figure}[b!]
\begin{center}
\includegraphics[width=0.5\textwidth]{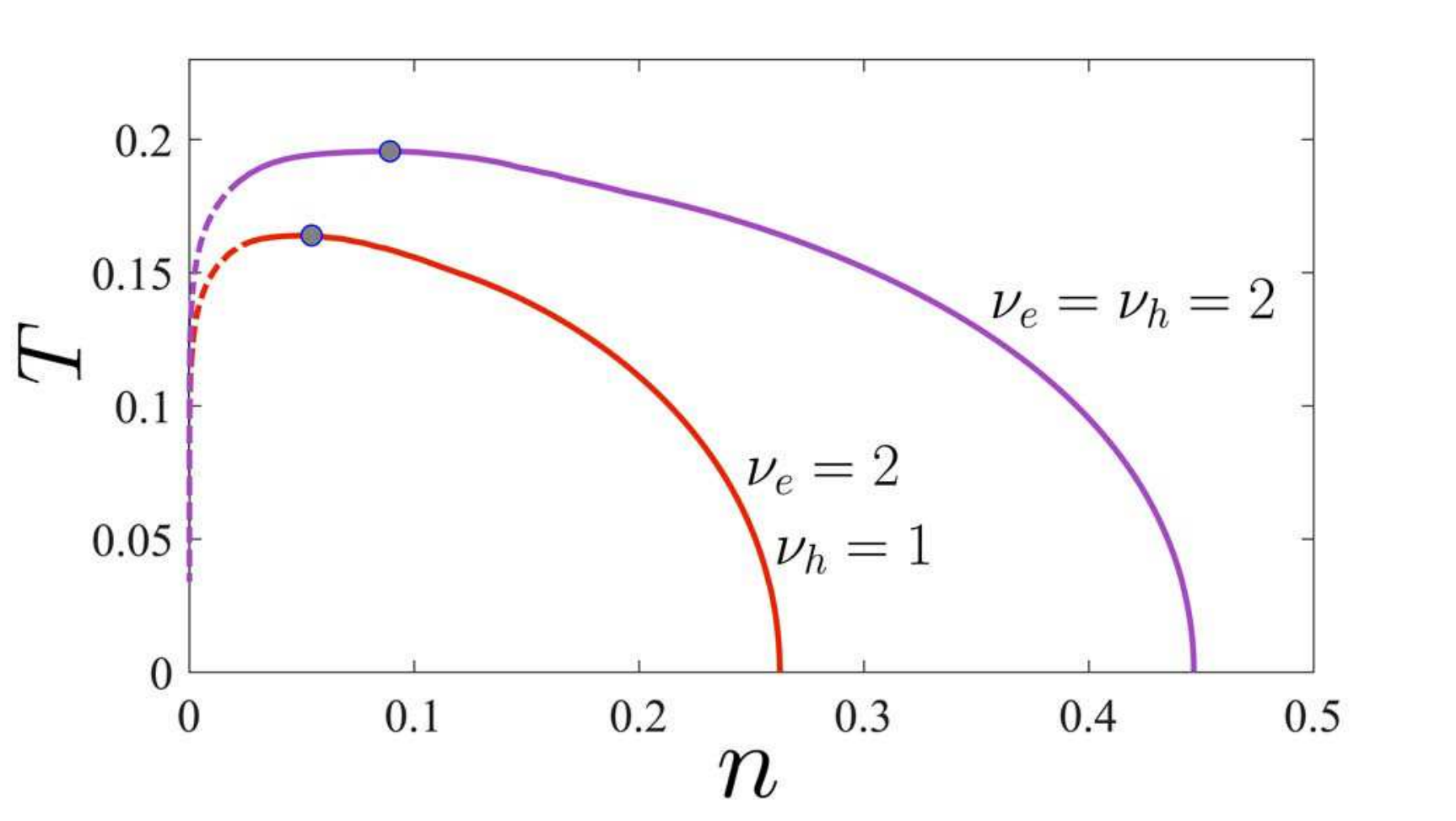}
\end{center}
\caption{\label{f5} Numerical calculation $n_\text{G}(T)$ and $n_\text{L}(T)$ for $\nu_e=2$ and $\nu_h=1$ (red curve) and $\nu_e=\nu_h=2$ (purple curve) for $\sigma=1$. The gray dots indicate the critical points of the gas--liquid transition. The dashed lines mark the parts of the curves in the region of low charge carrier densities, where RPA becomes inapplicable.}
\end{figure}

The set of pairs of points $n_\text{G}$ and $n_\text{L}$ forms a bell-shaped curve in the plane $(n,\,T)$. Two typical curves for a TMD monolayer with $\nu_e=2$ and $\nu_h=1$ and $\nu_e=\nu_h=2$ in the case of $\sigma=1$ are shown in Fig.~\ref{f5}.

For $\nu\gg1$, we find, with the same accuracy as \eqref{approx_zc}, the temperature dependences of the $e$-$h$ pair density in the gas and liquid phases and the chemical potential near the critical point $T_c-T\ll T_c$
\begin{equation}
\begin{split}
&n_c-n_\text{G}(T)=n_\text{L}(T)-n_c\\
&\approx\sqrt{\frac{6}{\gamma}}\left(1-\frac{1+s^2}{12\gamma}z^2_c\right)n_c\left(\frac{T_c-T}{T_c}\right)^{1/2},
\end{split}
\end{equation}
\begin{equation}
\mu(T)-\mu_c=\left\{2-\ln\left[(1-s^2)z^2_c\right]+\frac{1+s^2}{12}z^2_c\right\}\left(T_c-T\right).
\end{equation}

At low temperatures $T\ll T_c$, the density in the gas phase is exponentially small
\begin{equation}
n_\text{G}(T)\approx\frac{\nu T}{\pi\sqrt{1-s^2}}e^{-|\mu_0|/2T},
\end{equation}
where $\mu_0=\mu(n_\text{EHL},\,T=0)=E_\text{EHL}$. In the liquid phase, it deviates from the equilibrium EHL density quadratically in temperature
\begin{equation}
\begin{split}
&n_L(T)=n_\text{EHL}\left\{1-\frac{\pi^2}{6(1-s^2)}\left(\frac{1-\gamma}{2}\right)^{\frac{1+\gamma}{1-\gamma}}\right.\\
&\left.\times\left(1+\gamma\right)^{\frac{2}{1-\gamma}}z^{\frac{2\gamma}{1-\gamma}}_c\left(1+\frac{1+s^2}{6(1-\gamma)}z^2_c\right)\left(\frac{T}{T_c}\right)^2\right\}.
\end{split}
\end{equation}
The chemical potential correction is also quadratic in temperature
\begin{equation}\label{mu_lowT}
\begin{split}
&\mu(T)=\mu_0\left\{1+\frac{\pi^2\gamma}{6(1-s^2)}\left(\frac{1-\gamma}{2}\right)^{\frac{1+\gamma}{1-\gamma}}\right.\\
&\left.\times\left(1+\gamma\right)^{\frac{2}{1-\gamma}}z^{\frac{2\gamma}{1-\gamma}}_c\left(1+\frac{1+s^2}{6(1-\gamma)}z^2_c\right)\left(\frac{T}{T_c}\right)^2\right\}.
\end{split}
\end{equation}

The formulas \eqref{eq_zc}--\eqref{eq_zc_sigma1} and \eqref{approx_zc}--\eqref{mu_lowT} were written by us for an arbitrary $\gamma$ with the purpose that we will need these formulas when considering 3D layered systems, when $\gamma=\hspace{0.05cm}^1\hspace{-0.08cm}/_4$. The corresponding formulas are given in the work \cite{Andryushin1977a}.

A TMD monolayer, generally speaking, is a ``three-layer'' system in which a layer of transition metal atoms is inserted between two layers of chalcogen atoms. Therefore, it may turn out that the relevant value of the exponent in the correlation energy will be intermediate between $\hspace{0.05cm}^1\hspace{-0.08cm}/_4$ and $\hspace{0.05cm}^1\hspace{-0.08cm}/_3$.

In the case of a finite number of valleys, one can choose the effective exponent $\gamma$ in the expression for the correlation energy \eqref{Ecorr_An_gamma} so that simple approximate calculations agree with the exact ones. For this, two criteria can be proposed: 1) by the coincidence of $n_\text{EHL}/n_c$ with that obtained from the equation \eqref{n/nc}; 2) by coincidence of $|E_\text {EHL}|/T_c$ with that obtained from the equation \eqref{E/Tc}.

Based on the data in Table~\ref{t1}, we calculated for the case $\sigma=1$ the effective values of $\gamma$ and $A$ by both criteria: $\gamma_n$~and $A_n$ by the first criterion, $\gamma_T$ and $A_T$ by the second criterion (see Table~\ref{t2}).

\begin{table}[b!]
\caption{\label{t2} Values of the effective exponent and constants in the correlation energy.}
\begin{ruledtabular}
\begin{tabular}{cccccc}
$(\nu_e,~\nu_h)$ & (1,~1) & (2,~1) & (2,~2) & (4,~1) & (4,~4)\\
\hline
$\gamma_n$ & 0.341 & 0.320 & 0.305 & 0.291 & 0.281\\
$\gamma_T$ & 0.300 & 0.296 & 0.285 & 0.299 & 0.280\\
$A_n$ & 2.976 & 3.006 & 3.021 & 3.121 & 3.313\\
$A_T$ & 2.763 & 2.910 & 2.974 & 3.143 & 3.314\\
\end{tabular}
\end{ruledtabular}
\end{table}

The case of finite $\nu$ can be very different from the limiting case of $\nu\rightarrow\infty$. In a single-valley semiconductor, when the formula \eqref{Ecorr_An_gamma} for the correlation energy is unlikely to work, $\gamma_n>\hspace{0.05cm}^1\hspace{-0.08cm}/_3$. The coincidence of $\gamma_n$ and $\gamma_T$ improves as the number of valleys increases. For $\nu_e=\nu_h=4$ the number of valleys is large enough and $\gamma_n\approx\gamma_T\approx0.28$ and the question of choosing a criterion is insignificant. In this case, the effective exponent is closer to the layered $\gamma=\hspace{0.05cm}^1\hspace{-0.08cm}/_4$ than to $\hspace{0.05cm}^1\hspace{-0.08cm}/_3$, indicating an analogy between multi-valley and multi-layer systems.

In the case of a small number (one or two electron (hole) valleys), one should decide on the choice of the criterion. Numerical calculations of phase diagrams for systems with $\nu_e=2$, $\nu_h=1$, and $\nu_e=\nu_h=2$ using the expression \eqref{Ecorr_An_gamma} showed better agreement with the curves calculated earlier (see Fig.~\ref{f5}) for the effective exponent $\gamma_T$. Thus, in our opinion, the second criterion is more accurate.

The results obtained for the effective exponent $\gamma$ can be used to calculate the exchange-correlation contribution in the density functional method \cite{Kryachko2014, Jones2015, Bretonnet2017, InTech2019}.

\section{\label{s5}Metal--insulator transition in monolayer}

In the introduction, we pointed out a qualitative discrepancy between the theoretical work \cite{Rustagi2018} and the experiment \cite{Arp2019}. In the experiment, with increasing excitation, the metal--insulator transition preceded the gas--liquid transition, i.e. the metal--insulator transition occurred in the gas phase, and theoretical calculations predicted that this transition occurs in the liquid phase. We noted that the indicated discrepancy arose due to the unjustified use of the Keldysh potential in calculating the EHL ground state energy.

The authors of the work \cite{Rustagi2018} took as a criterion for the occurrence of a metal--insulator transition the equality $|\delta E_g|=E_x$, where $\delta E_g$ is the bandgap renormalization, $E_g=E^{(0)}_g+\delta E_g$ ($E^{(0)}_g$ is the bare value of the bandgap), $\delta E_g=\mu_\text{exch}+\mu_\text{corr}$. Qualitatively, this corresponds to the ``creeping'' of the bottom of the conduction band with an increase in the $e$-$h$ pair density to the exciton level, if its position is assumed to be unchanged. In our opinion, this criterion is incorrect because it does not take into account the screening of the Coulomb interaction at a high $e$-$h$ pair density. We are ``approaching'' the metal--insulator transition from the side of low densities. The renormalization of the bandgap in this density range can be neglected, because the density of free $e$-$h$ pairs in the insulator (exciton) phase is low.

We use vanishing the ground state energy (binding energy) of the exciton as a criterion for the metal--insulator transition
\begin{equation}
E_{exc}(n_{dm})=0.
\end{equation}
Here, $n_{dm}$ is the density of this transition.

To find the dependence $E_{exc}(n)$, we solve by the variational method the Schr\"{o}dinger equation in the momentum representation
\begin{equation}\label{Schrodinger}
\frac{p^2}{2m}\psi(\mathbf{p})-\int\frac{d^2q}{(2\pi)^2}V(\mathbf{q})\psi(\mathbf{q}-\mathbf{p})=E_{exc}\psi(\mathbf{p})
\end{equation}
($m$ is the reduced mass of an electron and a hole).

Screened Coulomb potential at a high $e$-$h$ pair density \cite{Andryushin1979, Silin1988, Andryushin1981}
\begin{equation}
V(\mathbf{q})=\frac{V_0(\mathbf{q})}{1+f(\mathbf{q})V_0(\mathbf{q})\Pi_0(\mathbf{q})},
\end{equation}
where the initial unscreened Coulomb interaction $V_0(\mathbf{q})$ is taken in the form of the Keldysh potential \eqref{keldysh_int}, $\Pi_0(\mathbf{q})$ is static 2D polarization operator of electrons and holes
\begin{equation}\label{Polarization_Operator}
\Pi_0(\mathbf{q})=\nu_e\Pi^e_0(\mathbf{q})+\nu_h\Pi^h_0(\mathbf{q}),
\end{equation}
\begin{equation*}
\Pi^{e,h}_0(\mathbf{q})=\frac{m_{e,h}}{\pi}\left\{1-\sqrt{1-\left(\frac{2q^{e,h}_F}{q}\right)^2}\theta\left(q-2q^{e,h}_F\right)\right\}.
\end{equation*}
Here, $\theta(x)$ is the Heaviside function,
\begin{equation*}
\theta(x)=\begin{cases}0,&\,x<0,\\1,&\,x>0.\end{cases}
\end{equation*}

The polarization operator is taken in the static limit, since we are interested in vanishing the exciton binding energy. $\Pi_0(\mathbf{q})$ differs from that used in the works \cite{Andryushin1979, Andryushin1981} in that the expression for $\Pi_0(\mathbf{q})$ takes into account the presence of several valleys ($\nu_e$ electron and $\nu_h$ hole ones).

The $f(\mathbf{q})$ function is the Hubbard correction to RPA taking into account the contribution of the exchange diagrams at high momentum transfers \cite{Andryushin1979, Andryushin1981, Hubbard1957}. In the case of an equal number of valleys $\nu_e=\nu_h=\nu\geq1$
\begin{equation}\label{General_Hubb_Correct1}
f(\mathbf{q})=1-\frac{1}{4\nu}\frac{q}{q+q_F}.
\end{equation}
Here, it is taken into account that the number of the exchange diagrams grows linearly in $\nu$, while the number of the loop diagrams grows quadratically in $\nu$. Therefore, the relative contribution of the former decreases as $1/\nu$.

In the case of $\nu_e\neq\nu_h$, the Fermi momenta of electrons and holes are different and it is convenient to take their geometric mean $\overline{q}_F=\sqrt{q^e_Fq^h_F}$, then
\begin{equation}\label{General_Hubb_Correct2}
f(\mathbf{q})=1-\frac{1}{2(\nu_e+\nu_h)}\frac{q}{q+\overline{q}_F}.
\end{equation}

The trial wave function is chosen in the form of the Fourier transform of an exponentially decreasing wave function of 2D exciton ($a$ is the variational parameter)
\begin{equation}\label{Psi_trial}
\psi(\mathbf{p})=\frac{\sqrt{8\pi}a^2}{(a^2+p^2)^{3/2}}.
\end{equation}

We checked the correctness of the variational calculations for the case of zero charge carrier density ($n=0$). Table~\ref{t3} summarizes the calculated and experimental data on the exciton binding energy for eight heterostructures.

\begin{table*}[t!]
\caption{\label{t3} Parameters of TMD monolayers, calculated ($|E_{exc}(0)|$) and experimental ($E_b$) exciton binding energies. $m_0$ is the free electron mass.}
\begin{ruledtabular}
\begin{tabular}{lcccccc}
\text{Heterostructure} & $m$ ($m_0$) & $r_0$ (\text{\AA}) & $\varepsilon_\text{eff}$ & $|E_{exc}(0)|$ (\text{meV}) & $E_b$ (\text{meV}) & \text{References} \\
\hline
\text{MoS}$_2$/\text{SiO}$_2$ & 0.320$\pm$0.04 & 16.926 & 2.45 & 336 & 310$\pm$40 & \cite{Eknapakul2014, Berkelbach2013, Rigosi2016}\\
$h$\text{BN}/\text{MoS}$_2$/$h$\text{BN} & 0.275$\pm$0.015 & 7.640 & 4.45 & 215 & 221$\pm$3 & \cite{Goryca2019}\\
$h$\text{BN}/\text{MoSe}$_2$/$h$\text{BN} & 0.350$\pm$0.015 & 8.864 & 4.4 & 226 & 231$\pm$3 & \cite{Goryca2019}\\
$h$\text{BN}/\text{MoTe}$_2$/$h$\text{BN} & 0.360$\pm$0.04 & 14.546 & 4.4 & 173 & 177$\pm$3 & \cite{Goryca2019}\\
\text{WS}$_2$/\text{SiO}$_2$ & 0.220 & 15.464 & 2.45 & 310 & 360$\pm$60 & \cite{Berkelbach2013, Rigosi2016, Rasmussen2015}\\
$h$\text{BN}/\text{WS}$_2$/$h$\text{BN} & 0.175$\pm$0.007 & 7.816 & 4.35 & 175 & 180$\pm$3 & \cite{Goryca2019}\\
\text{WSe}$_2$/\text{SiO}$_2$ & 0.230 & 18.414 & 2.45 & 283 & 370 & \cite{Berkelbach2013, Rasmussen2015, He2014}\\
$h$\text{BN}/\text{WSe}$_2$/$h$\text{BN} & 0.200$\pm$0.01 & 10 & 4.5 & 158 & 167$\pm$3 & \cite{Goryca2019}\\
\end{tabular}
\end{ruledtabular}
\end{table*}

To calculate the exciton binding energy in a MoS$_2$ monolayer on a SiO$_2$ substrate (MoS$_2$/SiO$_2$), the effective electron and hole masses were taken from the experiment \cite{Eknapakul2014}. The value of the screening length $r_0$ was taken from the theoretical work \cite{Berkelbach2013}. Our result was compared with the experimental exciton binding energy \cite{Rigosi2016}. A close value of the binding energy (340 meV) was obtained by the variational solution of the Schr\"{o}dinger equation in the coordinate space \cite{Ratnikov2020}.

For TMD monolayers encapsulated by thin layers of hexagonal boron nitride ($h$BN), the experimental data were taken from \cite{Goryca2019}.

\begin{figure}[b!]
\begin{center}
\includegraphics[width=0.49\textwidth]{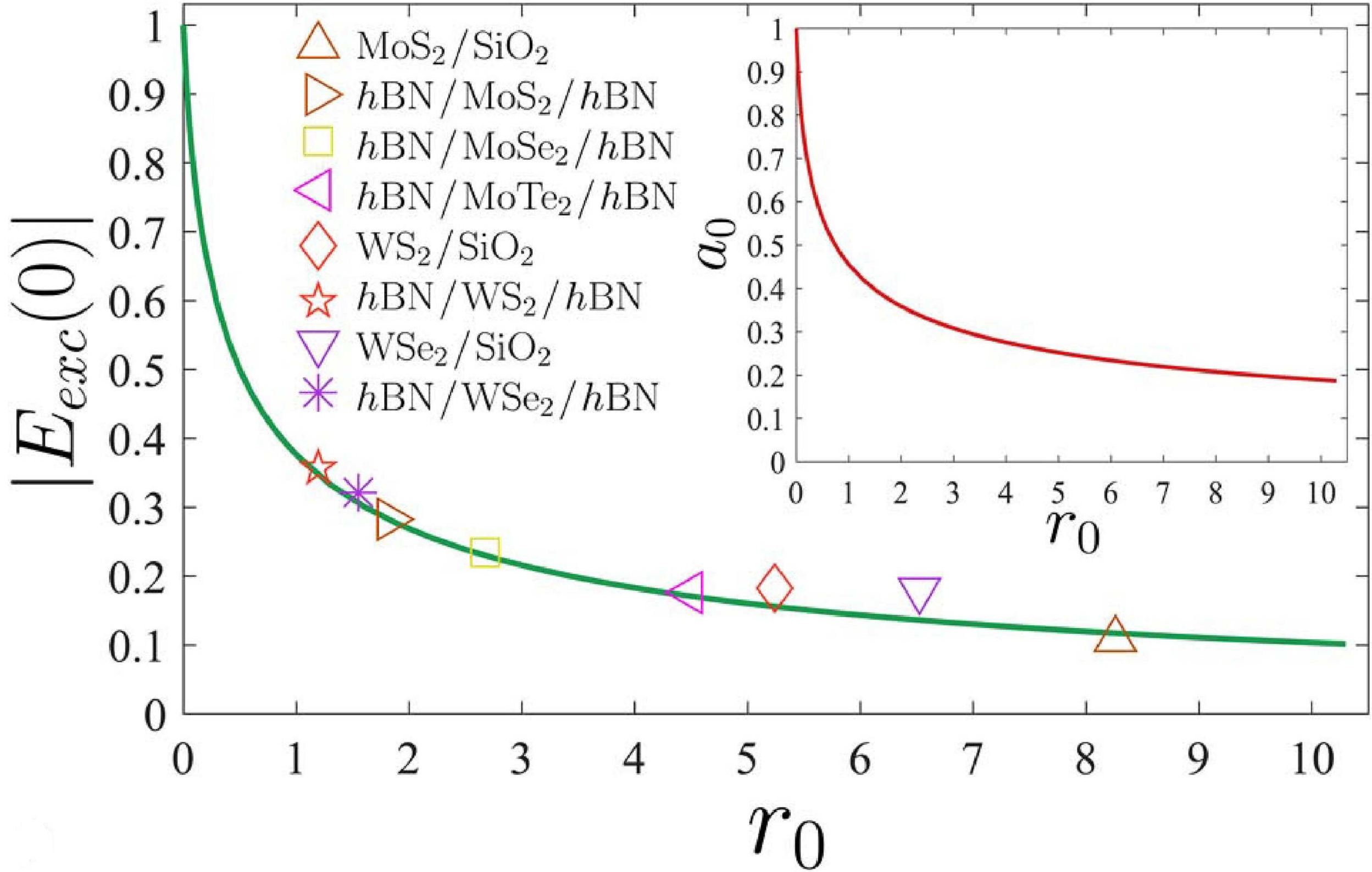}
\end{center}
\caption{\label{f6} Numerical calculation of the dependence of the exciton binding energy at zero charge carrier density $|E_{exc}(0)|$ on the screening length $r_0$. Points corresponding to several systems that were studied experimentally are highlighted. The inset shows the dependence of the variational parameter $a_0$ corresponding to the maximum binding energy on $r_0$.}
\end{figure}

For WS$_2$ and WSe$_2$ on a SiO$_2$ substrate (WS$_2$/SiO$_2$ and WSe$_2$/SiO$_2$), the calculated parameters were taken from the works \cite {Berkelbach2013, Rasmussen2015}. The experimental exciton binding energies in WS$_2$/SiO$_2$ and WSe$_2$/SiO$_2$ were obtained, respectively, in the works \cite{Rigosi2016} and \cite{He2014}.

For the MoS$_2$/SiO$_2$ and WS$_2$/SiO$_2$ heterostructures, calculated by us binding energies fall within the experimental error ranges. For other heterostructures, the calculated values are also in good agreement with the experimental ones, with the exception of WSe$_2$/SiO$_2$, for which the error range was not given \cite {He2014}. As can be seen from Table~\ref{t3}, on the whole, good agreement between our calculations and experiment was achieved.

\begin{figure}[b!]
\begin{center}
\includegraphics[width=0.5\textwidth]{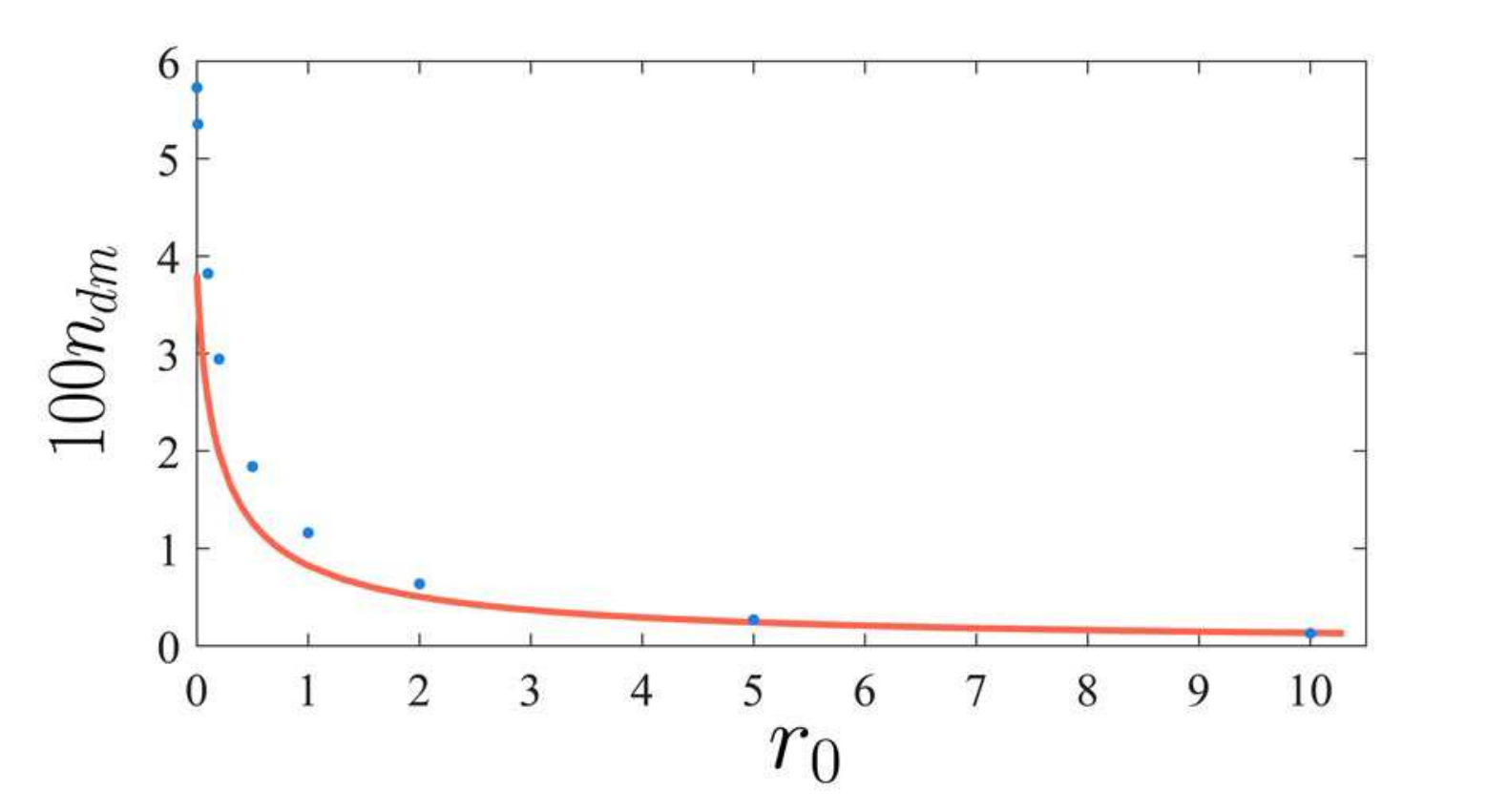}
\end{center}
\caption{\label{f7} Numerical calculation of the dependence of the metal--insulator transition density $n_{dm}$ on $r_0$ for a system with two electron and two hole valleys and equal masses of electrons and holes. The dots show the values of $n_{dm}$ from the work \cite{Andryushin1981} for a system with one electron and one hole valley and equal masses of electrons and holes.}
\end{figure}

We also calculated the dependence of the binding energy $|E_{exc}(0)|$ on $r_0$ (see Fig.~\ref{f6}). It should be noted that the error ranges for the dimensionless experimental values of the binding energy also appeared along the abscissa axis, since the unit of measurement $a_x$, by which we nondimensionalized $r_0$, has an error caused by inaccuracy in measuring the effective masses of charge carriers. For the same reason, inaccuracies in $E_x$ also introduce an additional error into the dimensionless value $E_{exc}(0)$. The experimental points for all heterostructures (except for WSe$_2$/SiO$_2$), taking into account the errors, coincide with the calculated curve. For the WSe$_2$/SiO$_2$ heterostructure, the authors of \cite{He2014} apparently calculated the exciton binding energy $E_b$ for a suspended sample in vacuum, which gave an overestimated value of $E_b$.

\begin{figure}[b!]
\begin{center}
\includegraphics[width=0.5\textwidth]{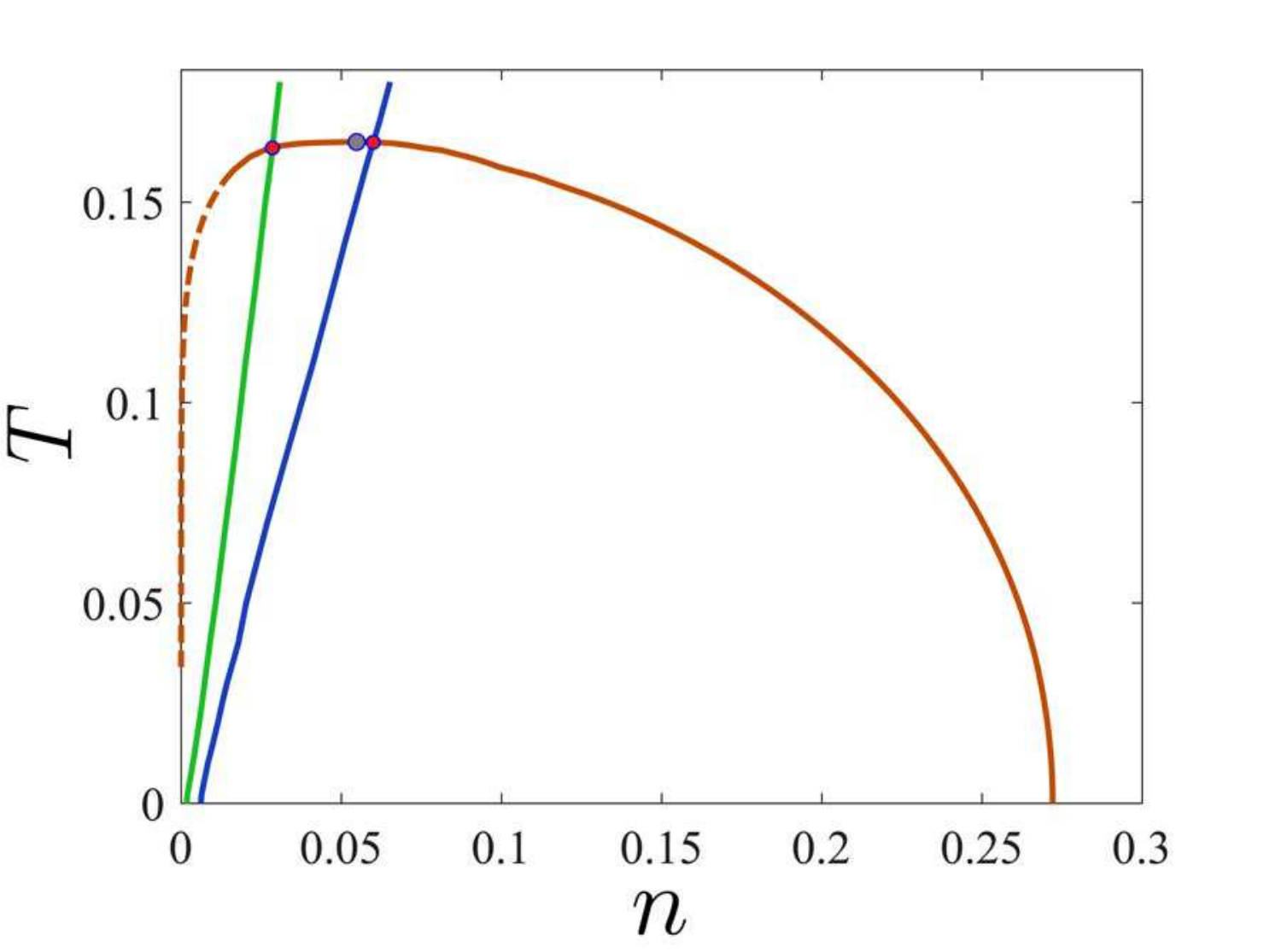}
\end{center}
\caption{\label{f8}  Temperature dependences of the metal--insulator transition density $n_{dm}(T)$ for heterostructures MoS$_2$/SiO$_2$ (green curve) and $h$BN/MoS$_2$/$h$BN (blue curve) and the gas--liquid transition curve (brown curve). The gray point indicates the critical point. The red dots indicate the intersection points of the $n_{dm}(T)$ curves with the gas--liquid transition curve.}
\end{figure}

We come to the conclusion that the method described above for the variational solution of the Schr\"{o}dinger equation is applicable to finding the exciton binding energy at zero $e$-$h$ pair density.

We have investigated the dependences of the metal--insulator transition density $n_{dm}$ at zero temperature on the screening length $r_0$ for various numbers of electron and hole valleys and the ratio of the charge carrier masses ($\sigma=0.8\div1.1$). The resulting curves $n_{dm}(r_0)$ for different $\nu_{e, h}$ and $\sigma$ turned out to be very close. We compared the results of our calculations for $\nu_e=\nu_h=2$ with those published in \cite{Andryushin1981} for $\nu_e=\nu_h=1$ (see Fig.~\ref{f7}). In the region of interest to us, $1\lesssim r_0\lesssim10$, the values of $n_{dm}$ are also very close to the calculated curve. This allows us to conclude that $n_{dm}$ is weakly dependent on $\nu_e$, $\nu_h$, and $\sigma$.

To calculate the dependence of the exciton binding energy on the $e$-$h$ pair density at $T\neq0$, we need a polarization operator at a finite temperature. After summation over discrete Matsubara frequencies and analytic continuation to real frequencies, we find its real part at zero frequency in the form
\begin{equation}\label{Polarization_Operator}
\Pi_0(\mathbf{q},\,T)=\nu_e\Pi^e_0(\mathbf{q},\,T)+\nu_h\Pi^h_0(\mathbf{q},\,T),
\end{equation}
\begin{equation}
\Pi^\alpha_0(\mathbf{q},\,T)=2\text{-}\text{-}\hspace{-0.357cm}\int\frac{d^2k}{(2\pi)^2}\frac{n^\alpha_F(\mathbf{k})-n^\alpha_F(\mathbf{k}+\mathbf{q})}
{\varepsilon^\alpha_{\mathbf{k}+\mathbf{q}}-\varepsilon^\alpha_\mathbf{k}},
\end{equation}
where $n^\alpha_F(\mathbf{p})=1/[\exp((\varepsilon^\alpha_\mathbf{p}-\mu^\alpha_\text{kin})/T)+1]$ is the Fermi distribution function of the $\alpha$th kind particles ($\alpha=e,\,h$) with the chemical potential $\mu^\alpha_\text{kin}$ (calculated for a gas of non-interacting particles), $\varepsilon^\alpha_\mathbf{p}=p^2/\eta_\alpha=\frac{1\pm s}{2}p^2$ is the dispersion law of the $\alpha$th kind particles (plus for electrons, minus for holes). The bar near the integral means that it is taken in the sense of the principal value.

Using the values of $m$, $r_0$, and $\varepsilon_\text {eff}$ given in Table~\ref{t3}, we calculated the temperature dependences of the metal--insulator transition density $n_{dm}(T)$ for MoS$_2$/SiO$_2$ and $h$BN/MoS$_2$/$h$BN heterostructures and compared them with the gas--liquid transition curve in Fig.~\ref{f8}. Points of intersection of curves on the phase diagram $(n,\,T)$ are $n_{dm}=0.029$ (2.8$\times10^{12}$ cm$^{-2}$) at $T_{dm}=0.164$ (638 K) for the first heterostructure and $n_{dm}=0.060$ (4.6$\times10^{12}$ cm$^{-2}$) at $T_{dm}=0.165$ (412 K) for the second heterostructure. The critical point of the gas--liquid transition $n_c=0.055$ (5.5$\times10^{12}$ cm$^{-2}$ and 4.2$\times10^{12}$ cm$^{-2}$ for the first and the second heterostructures, respectively) and $T_c=0.165$ (644 K and 412 K for the first and second heterostructures, respectively) is between them. Consequently, in MoS$_2$/SiO$_2$ the transition is in the gas phase, and in $h$BN/MoS$_2$/$h$BN the transition is in the liquid phase.

Note that the $n_c$ and $n_{dm}$ values for the $h$BN/MoS$_2$/$h$BN heterostructure are very close (the difference is $\lesssim10$\%) and we cannot strictly assert that the transition occurs in the liquid phase. Refinement of calculations of $n_{dm}$ using dielectric screening \cite{Lobaev1984} will be carried out by us in the next work.

\section{\label{s6}Conclusion}

In this work, we calculated the EHL phase diagram in monolayer heterostructures based on TMD for an arbitrary ratio of the electron and hole masses and the number of valleys.

We have demonstrated that in the multivalley case, the main contribution to the EHL formation is the correlation energy. We showed that the EHL binding energy and the EHL equilibrium density increase with the number of valleys in TMD. Note that for the same total number of valleys, the maximum values are achieved with the equal number of electron and hole valleys.

We have shown that the relationship between the critical temperature of the gas--liquid transition and the EHL binding energy $T_c\simeq0.1\left|E_\text{EHL}\right|$ is satisfied with good accuracy for various heterostructures.

The method developed by us made it possible to determine the power law of the dependence of the correlation energy on the density $E_\text{corr}=-An^\gamma$ and to estimate the exponent $\gamma$. Calculations have shown that it is limited to values for 3D (layered or multi-valley) and 2D cases, respectively, $\hspace{0.05cm}^1\hspace{-0.08cm}/_4<\gamma<\hspace{0.05cm}^1\hspace{-0.08cm}/_3$. An exception are single-valley TMDs, for which the power law for the correlation energy is inapplicable due to the absence of a small parameter.

We investigated the dependence of the exciton binding energy in monolayer TMD-based heterostructures on their parameters and found good agreement between the calculated and experimental values.

We studied the metal--insulator transition. As a criterion for such a transition, we used vanishing the exciton binding energy with an increase in the charge carrier density. The corresponding density $n_{dm}$ weakly depends on the number of valleys and the ratio of the electron and hole masses.

Usually, the metal--insulator transition is in the gas phase. We have shown that the dielectric environment of a TMD monolayer can determine in which phase (gas or liquid) the transition from the insulator to the metallic phase occurs. An example of such behavior are heterostructures based on the MoS$_2$ monolayer. If the monolayer is on a SiO$_2$ substrate, then the metal--insulator transition occurs in the gas phase, and if the monolayer is encapsulated by $h$BN, then the transition, apparently, occurs in the liquid phase. It should be noted that the point of the metal--insulator transition is very close to the critical point of the gas--liquid transition in an encapsulated monolayer and the obtained results are not very reliable and require refinement.

In our opinion, heterostructures based on a MoS$_2$ monolayer, freely suspended or located on a SiO$_2$ substrate, are the most interesting for studying EHL and the metal--insulator transition, since it has maximum $|E_\text{EHL}|$ and $n_\text{EHL}$.

It is interesting to investigate how the external dielectric environment and electron doping affect the metal--insulator transition and to find out in which phase (gas or liquid) it occurs.

\begin{acknowledgments}
P.~V.~Ratnikov thanks for financial support the Foundation for the Advancement of Theoretical Physics and Mathematics ``BASIS'' (the project no. 20-1-3-68-1).
\end{acknowledgments}

\end{document}